\documentclass{aa}  
\usepackage{graphicx}
\usepackage{txfonts}
\usepackage[version=3]{mhchem}

\usepackage{color}

\begin{document} 

   \title{PRODIGE - Envelope to Disk with NOEMA\thanks{Based on observations carried out under project number L19MB with the IRAM NOEMA Interferometer. IRAM is supported by INSU/CNRS (France), MPG (Germany) and IGN (Spain)}}
    
   \subtitle{II. Small-scale temperature structure and a streamer feeding the SVS13A protobinary using \ce{CH3CN} and DCN}

   \author{T.-H. Hsieh
          \inst{1}
          \and
            D. M. Segura-Cox\thanks{NSF Astronomy and Astrophysics Postdoctoral Fellow}
          \inst{2,1}
          \and
          J. E. Pineda
          \inst{1}
          \and
          P. Caselli
          \inst{1}
          \and
          L. Bouscasse
          \inst{3}
          \and
          R. Neri
          \inst{3}
          \and
          A. Lopez-Sepulcre
          \inst{3,4}
          \and
          M. T. Valdivia-Mena
          \inst{1}
          \and
          M. J. Maureira
          \inst{1}
          \and
          Th. Henning
          \inst{5}
          \and
          G. V. Smirnov-Pinchukov
          \inst{5}
          \and
          D. Semenov
          \inst{5}
          \and
          Th. M\"{o}ller
          \inst{6}
          \and
          N. Cunningham
          \inst{4}
          \and
          A. Fuente
          \inst{7}
          \and
          S. Marino
          \inst{8,9}
          \and
          A. Dutrey
          \inst{10}
          \and
          M. Tafalla
          \inst{7}
          \and
          E. Chapillon
          \inst{3,10}
          \and
          C. Ceccarelli
          \inst{4}
          \and
          B. Zhao
          \inst{11}
          }

   \institute{Max-Planck-Institut f\"{u}r extraterrestrische Physik, Giessenbachstrasse 1, D-85748 Garching, Germany\\
              \email{thhsieh@mpe.mpg.de}
         \and
            Department of Astronomy, The University of Texas at Austin, 2500 Speedway, Austin, TX, 78712, USA
        \and
            Institut de Radioastronomie Millim\'{e}trique (IRAM), 300 rue de la Piscine, F-38406, Saint-Martin d'H\`{e}res, France
        \and
            IPAG, Universit\'{e} Grenoble Alpes, CNRS, F-38000 Grenoble, France
        \and
            Max-Planck-Institut f\"{u}r Astronomie, K\"{o}nigstuhl 17, D-69117 Heidelberg, Germany
        \and
            I. Physikalisches Institut, Universität zu K\"{o}ln, Z\"{u}lpicher Str. 77, 50937 K\"{o}ln, Germany
        \and
            Observatorio Astron\'{o}mico Nacional (IGN), Alfonso XII 3, E-28014, Madrid, Spain
        \and    
            Jesus College, University of Cambridge, Jesus Lane, Cambridge CB5 8BL, UK
        \and
            Institute of Astronomy, University of Cambridge, Madingley Road, Cambridge CB3 0HA, UK
        \and
            Laboratoire d'Astrophysique de Bordeaux, Universit\'{e} de Bordeaux, CNRS, B18N, All\'{e}e Geoffroy Saint-Hilaire, F-33615 Pessac, France
        \and
            Department of Physics and Astronomy, McMaster University, Hamilton, ON L8S 4E8, Canada
             }

   \date{June 3, 2022}

 
  \abstract
   {}
   {We present high sensitivity and high-spectral resolution NOEMA observations of the Class 0/I binary system SVS13A, composed of the low-mass protostars VLA4A and VLA4B with a separation of $\sim$90 au. VLA4A is undergoing an accretion burst that enriches the chemistry of the surrounding gas. This gives us an excellent opportunity to probe the chemical and physical conditions as well as the accretion process.}
   {We observe the ($12_{\rm K}-11_{\rm K}$) lines of \ce{CH3CN} and \ce{CH3{}^13CN}, the DCN (3--2) line, and the \ce{C^18}O (2--1) line toward SVS13A using NOEMA.}
   {We find complex line profiles at disk scales which cannot be explained by a single component or pure Keplerian motion. By adopting two velocity components to model the complex line profiles, we find that the temperatures and densities are significantly different between these two components. This suggests that the physical conditions of the emitting gas traced via \ce{CH3CN} can change dramatically within the circumbinary disk. In addition, combining our observations of DCN (3-2) with previous ALMA high-angular-resolution observations, we find that the binary system (or VLA4A) might be fed by an infalling streamer from envelope scales ($\sim$700 au). If this is the case, this streamer contributes to the accretion of material onto the system with a rate of at least $1.4\times10^{-6}$ M$_\odot$ yr$^{-1}$.}
   {We conclude that the \ce{CH3CN} emission in SVS13A traces hot gas from a complex structure. This complexity might be affected by a streamer that is possibly infalling and funneling   material into the central region.}


   \maketitle
%
\section{Introduction} \label{sec:intro}
Complex Organic Molecules (COMs, C-bearing molecules containing $\geq$6 atoms, \citealt{he09}) have been discovered in high-mass star forming regions. 
It is believed that these COMs are sublimated from the icy mantles into the gas-phase in the hot ($\geq$100 K) and dense ($\geq10^7$ cm$^{-3}$) regions heated by the central source.
COMs are also present in chemically rich cores/disks (\citealt{ca12,va14}, and references therein).
In low-mass star forming regions, the COM emission is confined in a compact region with T$\geq$100 K surrounding the protostar and it is called ``hot corino'' \citep{ce04}.
In these environments, COMs are believed to be at least partially released from the icy mantles due to thermal evaporation and also formed in the gas phase upon sublimation of parent species.
However, it is also suggested that such COM emission can trace 
accretion shocks, disk atmospheres, or outflow cavities \citep{dr15,cs18,be20}.
It is therefore important to study in detail these objects and their surroundings to unveil their nature.

The symmetric top \ce{CH3CN} molecule, with its K-ladder transitions, provides a powerful tool to diagnose the temperature structure in hot dense regions.
It is commonly observed in high-mass star forming regions \citep{ga10,be11,hu14,il16}.
In low-mass star forming regions, \ce{CH3CN} has been detected in NGC1333 IRAS4B/2A \citep{bo07}, IRAS16293-2422 \citep{bi08,ca18}, several sources from the CALYPSO survey \citep{be20} and PEACHES survey \citep{ya21}.
In addition, with high-angular-resolution observations, \ce{CH3CN} has been found to directly trace the Class II disk surfaces in recent works \citep{be18,lo18}.

SVS13  is a multiple system located in the NGC1333 cluster within the Perseus molecular cloud complex ($d=$293 pc, \citealt{or18}).
The multiple system includes a Class I close binary SVS13A (VLA 4A/B or Per-emb-44B/A) with a separation of $0\farcs3$ ($\sim$90 au) and three companions at a distance of 5\farcs3 (SVS13A2, $\sim$1600 au), 14\farcs9 (SVS13B, $\sim$4400 au), 34\farcs5 (SVS13C, $\sim$10,000 au) based on continuum emission \citep{an04,to16,to18,se18,ty20}. 
The close binary drives multiple jet/outflows \citep{le17,le16,st18}, and 
the presence of jet wiggling suggests that an extra undetected source may be present ($\sim20-30$ au from VLA4B, \citealt{le17}). A dusty tail/spiral at 1.3 mm is seen connected to VLA4A \citep{to18}, and it is suggested to be a fragmenting disk due to gravitational instability.
Recently, with the Atacama Large Millimeter Array (ALMA) high-angular-resolution ($0\farcs1$ or $\sim$30 au) 0.9 mm observations, \citet{di21} find a circumbinary disk, as well as two circumstellar disks around VLA4A/B. 
They estimate the stellar masses using orbital motions (see also \citealt{ma20} for a similar method on the Class 0 binary IRAS 16293A).
The total mass of the binary is found to be 1.0$\pm$0.4 $M_\odot$, and the masses of the individual protostars are estimated to be 0.27$\pm$0.10 $M_\odot$ for VLA4A and 0.60$\pm$0.20 $M_\odot$ for VLA4B.
SVS13A is also considered as a protostar system undergoing an accretion burst given the high bolometric luminosity, $L_{\rm bol}=45.3~L_\odot$ \citep{hs19}.
The origin of such accretion outbursts are still unclear \citep{au14}. They can be triggered by gravitational instability \citep{vo10} or magnetorotational instability \citep{ar01}. 
In addition, anisotropic envelope accretion could also lead to bursts. Indeed, recent observations find infalling streamers at envelope scales,
which are also suggested to change the protostellar accretion by funneling material into the inner region \citep{pi20,al20,gi21,mu21,ca21,ga22,th22,va22,Pineda2022-PP7}.

Regarding chemical complexity, a large number of COMs are detected toward SVS13A
\citep{de17,be20,ya21}.
Water emission is detected with a broad linewidth $\sim5-20$ km s$^{-1}$ (NGC1333-IRAS3A; \citealt{kr12}).
Using the IRAM 30m telescope, \citet{co16} detected deuterated water (HDO) emission toward SVS13A, and a non-local thermal equilibrium (non-LTE) analysis suggests that it comes from a hot (150$-$260 K), dense ($\geq3\times10^7$ cm$^{-3}$), and compact ($R\sim25$ au) region. 
Given such physical conditions, they suggest that the COM emission comes from a hot corino inside SVS13A.
\citet{di21} find a circumstellar disk around VLA4A with a radius of $\sim30$ au traced by Ethylene Glycol (aGg'-\ce{(CH2OH)2}).
Furthermore, \citet{bi17,bi19} found that the COM abundances are not significantly different between this Class I source and Class 0 sources, but the deuteration fraction of organic molecules is lower.
Later, a high \ce{CH2DCN}/\ce{CH3CN} ratio is measured in SVS13A, suggesting that the \ce{CH3CN} was formed via gas-phase reactions during the pre-stellar phase \citep{bi22a}.

In this work, we present NOrthern Extended Millimeter Array (NOEMA) \ce{CH3CN}, \ce{CH3{}^13CN}, DCN, and \ce{C^18O} observations toward SVS13A.
Thanks to the wide bandwidth and high spectral resolution achievable with the PolyFiX correlator, along with the sensitivity achievable with NOEMA,
we are able to study the temperature and kinematics of SVS13A through \ce{CH3CN} and DCN.
In Section \ref{sec:obs}, we present the data and observations used in this paper. The results and analysis are detailed in Section \ref{sec:res}. In Section \ref{sec:dis}, we discuss the scientific outputs, and the conclusions are summarized in Section \ref{sec:con}.

\section{Observations}
\label{sec:obs}
\subsection{NOEMA observations}
We observed SVS13A using the NOrthern Extended Millimeter Array (NOEMA) of the Institut de Radioastronomie Millim\'{e}trique (IRAM).
The observations are part of the 
MPG - IRAM observing program PROtostars \& DIsks: Global Evolution (PRODIGE) (Project ID: L19MB002, PIs: P. Caselli and T. Henning).
The observations consist of C-array configurations executed on 2019 December 29th and 2020 January 5th, as well as D-array configurations executed on 2020 August 6th and 2020 September 7th.
In all executions, ten antennas were used.
The on-source observing time is 2.74 hours for the C-array and 2.22 hours for the D-array.
The bandpass calibrators were 3C84 and 3C454.3. 
Two phase calibrators, 0333+321 and 0322+222, were observed for all executions.
The flux calibrators were LKHA101 and MWC349.
Combining the C- and D-arrays with baselines 15.6 to 368 m (11.5 to 270.0 k$\lambda$)
resulted in a synthesized beam size of $1\farcs2\times0\farcs7$ with natural weighting for line cubes and $1\farcs1\times0\farcs6$ with uniform weighting for the continuum image at $\sim$220 GHz. The maximum
recoverable scale is $22\farcs07$.

The spectral setup consists of four broad-band low-resolution windows and 39 narrow-band high-resolution windows within the broad-band frequency region. The broad-band windows cover the frequency ranges of 214.7-218.8 GHz, 218.8-222.8 GHz, 230.2-234.2 GHz, and 234.2-238.3 GHz with a channel width of 2 MHz ($\sim2.7$ km s$^{-1}$).
In this paper, we present three narrow-band windows that were designed to cover the CH$_3$CN $J=12-11$ K-ladder, DCN $J=3-2$ and \ce{C^{18}O} $J=2-1$. The channel width of these three windows is 62.5 kHz ($\sim$0.08 km s$^{-1}$).

The data reduction was performed by the NOEMA pipeline calibration routine using GILDAS-CLIC package\footnote{https://www.iram.fr/IRAMFR/GILDAS}, with additional manual flagging.
We performed phase-only self-calibration with the GILDAS {\tt MAPPING} package using the continuum maps from the broad-band spectral windows (see section \ref{sec:cont}). The iterative self-calibration was performed with  solution intervals of 300 s, 135 s, and 45 s on the data which includes only line-free channels.

For imaging of the line emission, continuum subtraction for the narrow spectral windows is done using the {\tt uv\_baseline} task in GILDAS/{\tt MAPPING} package. Given the line-rich spectra, the line free regions are manually selected after visually examining the spectra. Imaging is done using the {\tt clean} task with natural weighting for line cubes in order to increase the sensitivity. 
The continuum map was produced with robust=1 Briggs weighting
because the sensitivity is already high given the broadband width (4 GHz for each). With a manual clean mask, multi-scale clean algorithm is performed with a clean threshold of 1$\sigma$. The resulting data cubes have a beam size of $1\farcs23\times0\farcs72$ with a pixel size of $0\farcs15$.
The rms noise level is $\sim0.013$ Jy beam$^{-1}$ for these image cubes with a channel width of $\sim0.08$ km s$^{-1}$.
In comparison, the beam size of the continuum image is $1\farcs07\times0\farcs61$ with a pixel size of $0\farcs15$, and the rms noise level is 0.5 mJy beam$^{-1}$ (0.02 K).

\subsection{ALMA archival data}
We use the ALMA archival continuum and \ce{C^18O} maps from VANDAM\footnote{https://dataverse.harvard.edu/dataverse/VANDAM} \citep{to16,to18}. 
The ALMA project ID is 2013.1.00031.S.
The observations and data calibration are detailed in \citet{to18}. The angular resolutions are $0\farcs26\times0\farcs16$ for continuum image and $0\farcs36\times0\farcs26$ for \ce{C^18O} channel maps given the weighing and uv-range in use (see \citealt{to18}). The channel width of the \ce{C^18O} maps is 0.25 km s$^{-1}$.

\begin{figure*}
\includegraphics[width=1\textwidth]{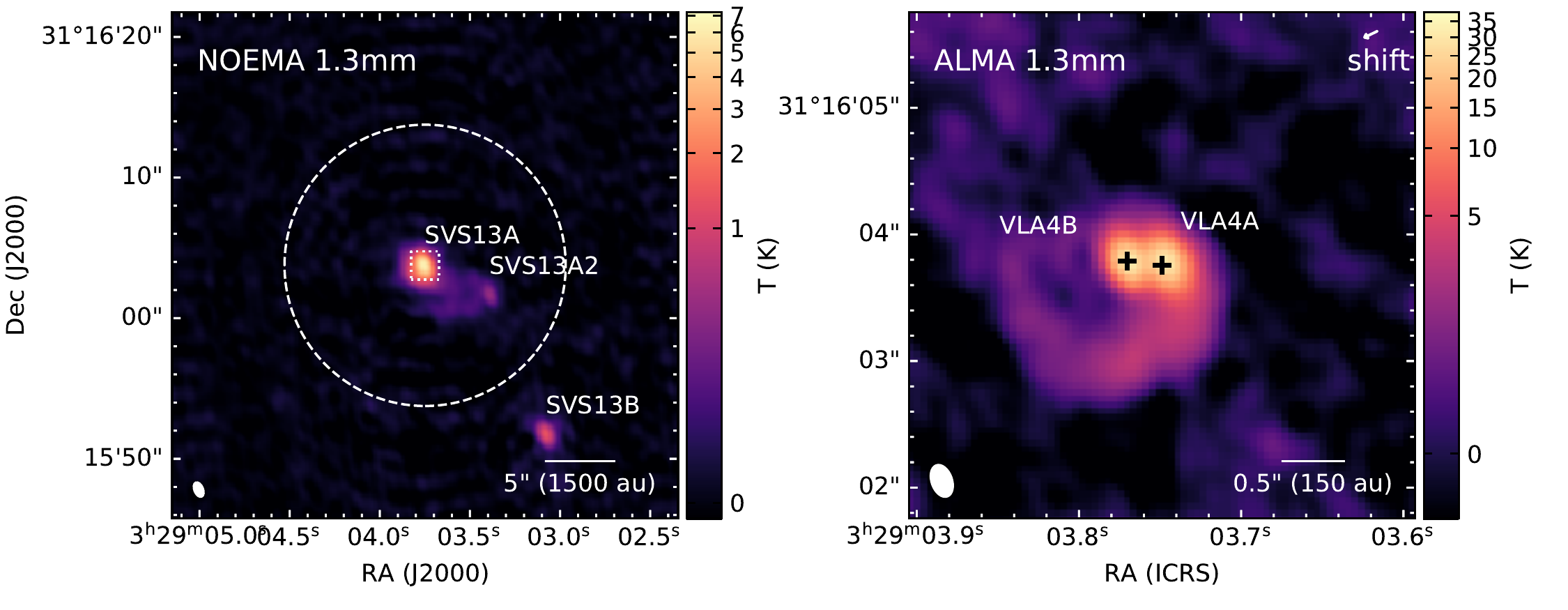}
\caption{1.3 mm continuum emission toward the SVS13 region from NOEMA (left, rms$\sim$20 mK) and ALMA (right) observations. In the left panel, the dashed circle shows the primary beam of the NOEMA observation and the dashed box represents the FoV in the right panel. The black plus signs represent the positions of VLA4A/B from \citet{hs19}. The arrow in the top right corner indicates the shift ($0\farcs13$) of the self-calibrated ALMA image.
}
\label{fig:cont}
\end{figure*}

As the ALMA data from \citet{to18} were self-calibrated, we need to apply a shift to the coordinates to match the position of VLA4A/B as observed in ALMA observations by \citet{hs19} for which no self-calibration was performed. The spatial resolutions of these observations are $0\farcs27$ and $0\farcs11$, respectively. 
The final shift applied to the ALMA observation is $\sim0\farcs13$ as shown in Figure \ref{fig:cont}.

\section{Results and Analysis}
\label{sec:res}
\subsection{Continuum images}\label{sec:cont}
The left panel of Figure \ref{fig:cont} shows the NOEMA continuum image.
The multiple system consisting of SVS13A, SVS13A2, and SVS13B is resolved. 
Extended dust emission is also detected between SVS13A and SVS13A2, separated by $\sim$1600 au.
The right panel shows a zoom-in image to SVS13A from the high-resolution ALMA observation at 1.3 mm from VANDAM \citep{to18}.
The close binary is resolved, as well as a spiral pattern that is interpreted as a fragmenting disk due to gravitational instability \citep{to18}.


\subsection{\ce{CH3CN} and \ce{CH3{}^13CN} emission} \label{sec:ch3cn}
The high-spectral resolution narrow-band spectrum of \ce{CH3CN} and \ce{CH3{}^13CN} toward the peak continuum emission of SVS13A is shown in Figure 2.
The \ce{CH3CN} J=12--11 K-ladder emission (K=0 -- K=7) is clearly detected and, although with lower S/N ratios, the \ce{CH3{}^13CN} J=12--11 K=0 to K=5 emission is also marginally detected (S/N$\sim$3.1 for K=5). A number of additional lines are identified using an automatic line identification algorithm (XCLASS, \citealt{mo17}), applied to the full broad-band windows with a spectral coverage of 16 GHz. 
LTE radiative transfer models for each identified molecule are provided. 
In order to check the line contamination, the identified lines with Einstein coefficients $>10^{-6}$ s$^{-1}$ and $E_{\rm up}<500$ K are labeled.
Based on the LTE model, the emission from the \ce{CH3OCHO}
(A$_{\rm ul}=6\times10^{-6}$ and $E_{\rm up}=382$ K) and \ce{^33SO2} (A$_{\rm ul}=1\times10^{-4}$ and $E_{\rm up}=50$ K) labeled in purple is very weak or undetectable and should not significantly affect the later \ce{CH3{}^13CN} analysis. However, a detailed study including full high-spectral resolution windows is required.
We will present the full chemical inventory in a future paper (Hsieh et al., in prep.).

Figure \ref{fig:uv_chan} (left and center) shows the integrated intensity of the \ce{CH3CN} (12$_3$-11$_3$) and \ce{CH3^13CN} (12$_1$-11$_1$) emission, the brightest component with less contamination from other line emission. 
As the emission appears compact, we fit it with a Gaussian in uv-space.
The derived size of the \ce{CH3CN} emission is $0\farcs42\pm0\farcs02\times0\farcs34\pm0\farcs01$ ($\sim$110 au), likely overestimated due to insufficient resolution.
The size of \ce{CH3{}^13CN} is found to be 0.3$\arcsec$$\times10^{-5}$$\arcsec$ from the fitting for which the minor axis of $10^{-5}$ arcsec is unlikely constrained.
Indeed, higher-resolution observations (beam$\sim0\farcs5$) derived a smaller size of $0\farcs3$ (90 au) for the COM emitting region \citep{be20}, likely also an upper limit (see section \ref{sec:cassis}).

\begin{figure*}
\includegraphics[width=1\textwidth]{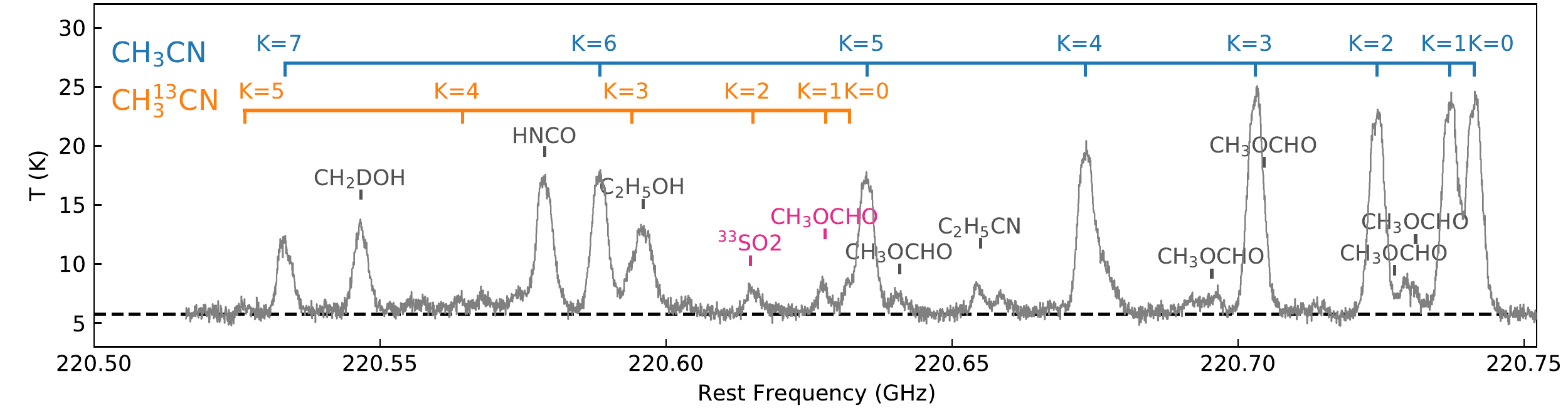}
\caption{Spectrum of \ce{CH3CN} and \ce{CH3{}^13CN} toward the peak of the continuum emission. The labels of the line frequencies, blue (\ce{CH3CN}), orange (\ce{CH3{}^13CN}), grey and pink bars, are shifted assuming a systemic velocity of 8.1 km s$^{-1}$. 
We only mark lines with an Einstein coefficient $>10^{-6}$ s$^{-1}$ and $E_{\rm up}<500$ K; 
The Einstein coefficients are $6\times10^{-6}$ s$^{-1}$ ($E_{\rm up}=382$ K) for the \ce{CH3OCHO} line and $1\times10^{-4}$ s$^{-1}$ ($E_{\rm up}=50$ K) for the \ce{^33SO2} line in pink while both are very weak or undetectable based on the XCLASS model from wideband spectra including more transitions.
The horizontal dashed line indicates the continuum level of 5.8 K determined from line free channels. The sensitivity per channel is 380 mK.
}
\label{fig:spec}
\end{figure*}

\subsection{CASSIS fitting}
\label{sec:cassis}
The \ce{CH3CN} K-ladder gives us the opportunity to study the physical conditions of the hot gas. We perform the CASSIS\footnote{http://cassis.irap.omp.eu/} \citep{va15}  Markov-chain Monte Carlo (MCMC) fitting to the \ce{CH3CN} (J=12--11, K=0--7) and \ce{CH_{3}{}^{13}CN} (J=12--11, K=0--5) emission.
The line parameters are taken from CDMS\footnote{https://cdms.astro.uni-koeln.de/cdms/portal/} \citep{mu05,en16};
the parameters for \ce{CH3CN} and \ce{CH3{}^13CN} are originally provided by \citet{mu15} and \citet{mu09} and references therein.
The partition function Q(T) of \ce{CH3CN} includes vibrational contribution up to v8=1, while only the ground state is available for \ce{CH3{}^13CN} in the CDMS database. 
Vibrational correction of the isotopologue \ce{CH3{}^13CN} can be done by scaling Q(T) using the ratios of the Q(T) of \ce{CH3CN} with and without vibrational contribution.
Since the ground state Q(T) of \ce{CH3CN} and that of \ce{CH3{}^13CN} are almost the same (difference $<$0.06 \% at T$<$500 K), the scaled partition function of \ce{CH3{}^13CN}, i.e., vibrational corrected, will be the same as that of the \ce{CH3CN} one within a factor of 0.06 \%.
We thus use the partition function of \ce{CH3CN} including vibrational contributions for the \ce{CH3{}^13CN} in our MCMC fitting.
We excluded the \ce{CH3CN} K=4 component, \ce{CH_{3}{}^{13}CN} K=3 component, and the blue-shifted wing of the \ce{CH3CN} K=6 component from our fitting due to contamination from nearby lines (see Figure \ref{fig:spec}).
A continuum level of 5.77 K is determined based on the line free channels (Figure \ref{fig:spec}) and is taken as an input for the simultaneous line fit performed by CASSIS.
The technical information about the MCMC sampling is detailed in Appendix \ref{sec:mcmc}.

Figure \ref{fig:cassis} (top panel) shows the best-fit spectrum with a one-component LTE model.
Table \ref{tab:cassis} lists best-fit column density, rotation temperature, linewidth, source size and central velocity, and isotopologue ratio (\ce{CH3CN}/\ce{CH3{}^13CN}). The posterior probability distribution of each parameter is shown in Figure \ref{fig:corner_lte}.
The resulting \ce{CH3CN} column density and temperature are $5.13\times10^{16}$ cm$^{-2}$ and 247.8 K, respectively, with a source size of $0\farcs27$. The column density and temperature are larger by a factor of 1.3--8.5 and 1.4--1.7 than that in \citet{be20} and \citet{ya21} (Table \ref{tab:cassis}).
The difference between our LTE model and the literature most likely comes from the assumption of the source size in the previous works.
In \citet{be20}, a source size of $0\farcs3$ based on Gaussian fitting in uv-space was adopted, and \citet{ya21} fixed a size of $0\farcs5$ for the unresolved case.
As shown in Figure \ref{fig:corner_lte}, the column density and excitation temperature are highly correlated with the source size; we note here that the optical depths from the fitting are 0.36--3.20 for the \ce{CH3CN} J=12--11 K=0--7 emission. 
It is clear that the adopted source size will determine the column density and excitation temperature as well as the optical depth $\tau$, although 
this may not significantly change the relative abundances of COMs in the literature \citep{be20,ya21}. In our case, with the source size as a free parameter, the optical depth is determined based on the line intensity ratios.
Because the K=3n (i.e., 3 and 6) transitions have a degeneracy larger than the others by a factor of 2, increasing the column density will change their optical depths much more than the others (i.e., the K=3n components will saturate faster than their neighboring transitions).
The opacities of the best-fit model are listed in Table \ref{tab:tau}.
In this case, the source size scales the intensities of all transitions, and the temperature affects the level populations as well as the optical depths.
As a result, the high sensitivity data enable us to determine the optical depths as well as column density from the accurate measurement of the intensity ratios.

This one-component model does not reproduce well the peak of the optically thick transitions (K=0--3). Especially for the K=3 component, the synthetic line profile is much broader than that from the observation (see also Figure \ref{fig:asym}). 
This suggests that the optical depths are not properly estimated.
Interestingly, the isotopologue ratio [\ce{CH3CN}/\ce{CH3{}^13CN}]$\simeq16$ is surprisingly low compared to the canonical ratio \ce{^{12}C}/\ce{^{13}C}$\simeq68$ \citep{mi05}. 
A possibility is that the optical depth, $\tau$, is not properly estimated by this simple 1-component model.
We therefore perform a fitting that takes \ce{CH3CN} and \ce{CH3{}^13CN} as independent species with different temperatures, velocities and linewidths (Figure \ref{fig:cassis}, 2nd panel). This gives us very different temperatures from \ce{CH3CN} (246.6 K) and \ce{CH3{}^13CN} (166.8 K), implying that they originate from different regions.
In addition, the observed line profile shows an asymmetric structure; the emission is stronger in the blue shifted side in the low-energy transitions, but reversed in the K=7 component (Figure \ref{fig:asym}).

It is most likely that a one-component model cannot well fit the \ce{CH3CN} emission from this known complex structure.
We thus perform a two-component LTE model to fit the observational spectrum, which results in a better fit (Figure \ref{fig:cassis}, 3rd panel); the synthetic spectral profile and observed profile have similar emission peaks in the K=0 and K=1 components and similar width in the K=3 component. 
The best-fit parameters are listed in Table \ref{tab:cassis} and their opacities are listed in Table \ref{tab:tau}.
These two components are assumed to be spatially separated, i.e., the interaction mode is turned off in our CASSIS setup. 
To resolve such a complicated scenario requires data with a higher spatial resolution.
In fact, the binary system has been spatially resolved by continuum emission \citep{to18} and velocity gradient toward SVS13A is revealed via other COM emission \citep{di21}.
A compact optically-thick red-shifted component and a thin blue-shifted component are separated, which agree with the asymmetric structures in the line profiles.
This result suggests that the traditional rotation diagram or one component model cannot accurately describe this complex profile. In this case, the two-component model implies the presence of an optically thick component. This would result in an underestimation of the column density when using one-component fitting.
It is however noteworthy that although the isotopologue ratio is higher in the two-component model compared with the one-component model, they are still lower than the canonical ratio (Table \ref{tab:cassis}).

Further, we estimate the H$_2$ density by applying the non-LTE radiative transfer code RADEX \citep{va07} in CASSIS (Figure \ref{fig:cassis}, bottom panel).
The collision rates are taken from LAMDA\footnote{https://home.strw.leidenuniv.nl/~moldata/} as calculated by \citet{gr86}. We note here the geometry of the emitting gas is described as a sphere in the RADEX calculation and the two components are spatially separated. We have a compact optically thick component and a relatively extended optically thin component at the blue- and red-shifted sides, similar to the LTE case (Table \ref{tab:cassis}). The compact component is close to LTE with $n_{\rm H_2}\sim1.2\times10^8$ cm$^{-3}$ (critical density $\sim10^7-10^8$ cm$^{-3}$ for a temperature of $\sim$200 K). Given the $n_{\rm H_2}$ densities of 6.2$\times$10$^{6}$ cm$^{-3}$ and $>$1.2$\times$10$^{8}$ cm$^{-3}$, we estimate the enclosed masses of 1.1$\times$10$^{-5}$ M$_\odot$ within 0\farcs287 (r=42 au) for the extended component and $>$2.4$\times$10$^{-5}$ M$_\odot$ within 0\farcs141 (r=21 au) for the compact component. These values are smaller than the disk masses (0.004--0.009 $M_\odot$ for VLA4A and 0.009--0.030 $M_\odot$ for VLA4B) derived from the dust continuum emission (\citealt{di21}). A possibility is that the optically thick \ce{CH3CN} emission does not trace the dense disk midplanes. Alternatively, the $n_{\rm H_2}$ density can be higher than the critical density by $\sim$2 order of magnitudes.
Noteworthy is that here the comparison simply assumes a uniform density to give a very rough idea.

\begin{figure*}
\begin{center}
\includegraphics[width=0.95\textwidth]{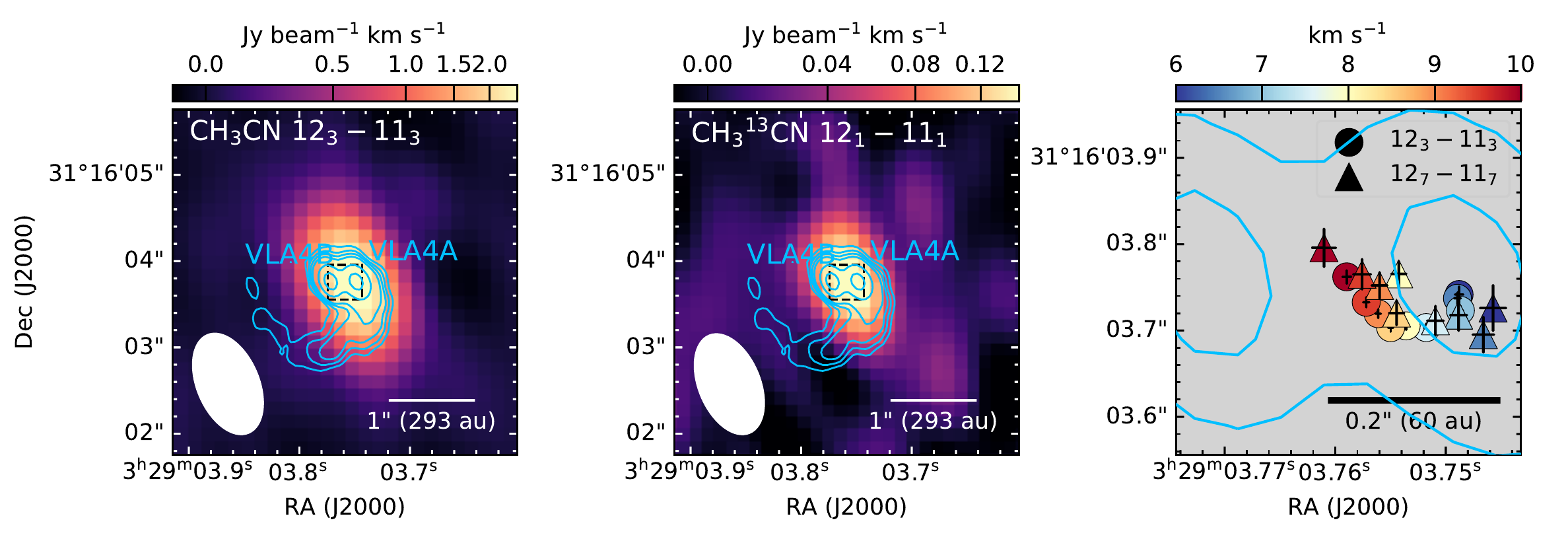}
\end{center}
\caption{({\it Left}) Integrated intensity map of \ce{CH3CN} 12$_3$-11$_3$ emission (4.0-12.0 km s$^{-1}$). The contours represent the ALMA 1.3 mm continuum emission from Figure \ref{fig:cont}. with levels of 3$\sigma$, 5$\sigma$, 7$\sigma$, 10$\sigma$, 30$\sigma$, and 70$\sigma$. The dashed box shows the field of view in the right panel.
({\it Medium}) Same as left but for \ce{CH3{}^13CN} 12$_1$-11$_1$ emission.
({\it Right}) Gaussian centers from the uv-domain fitting.  The colored circles (\ce{CH3CN} $12_3-11_3$) and triangles (\ce{CH3CN} $12_7-11_7$) show the positions from the uv-domain fitting with the color indicating the velocity.
}
\label{fig:uv_chan}
\end{figure*}

\begin{figure*}
\begin{center}
\includegraphics[width=0.99\textwidth]{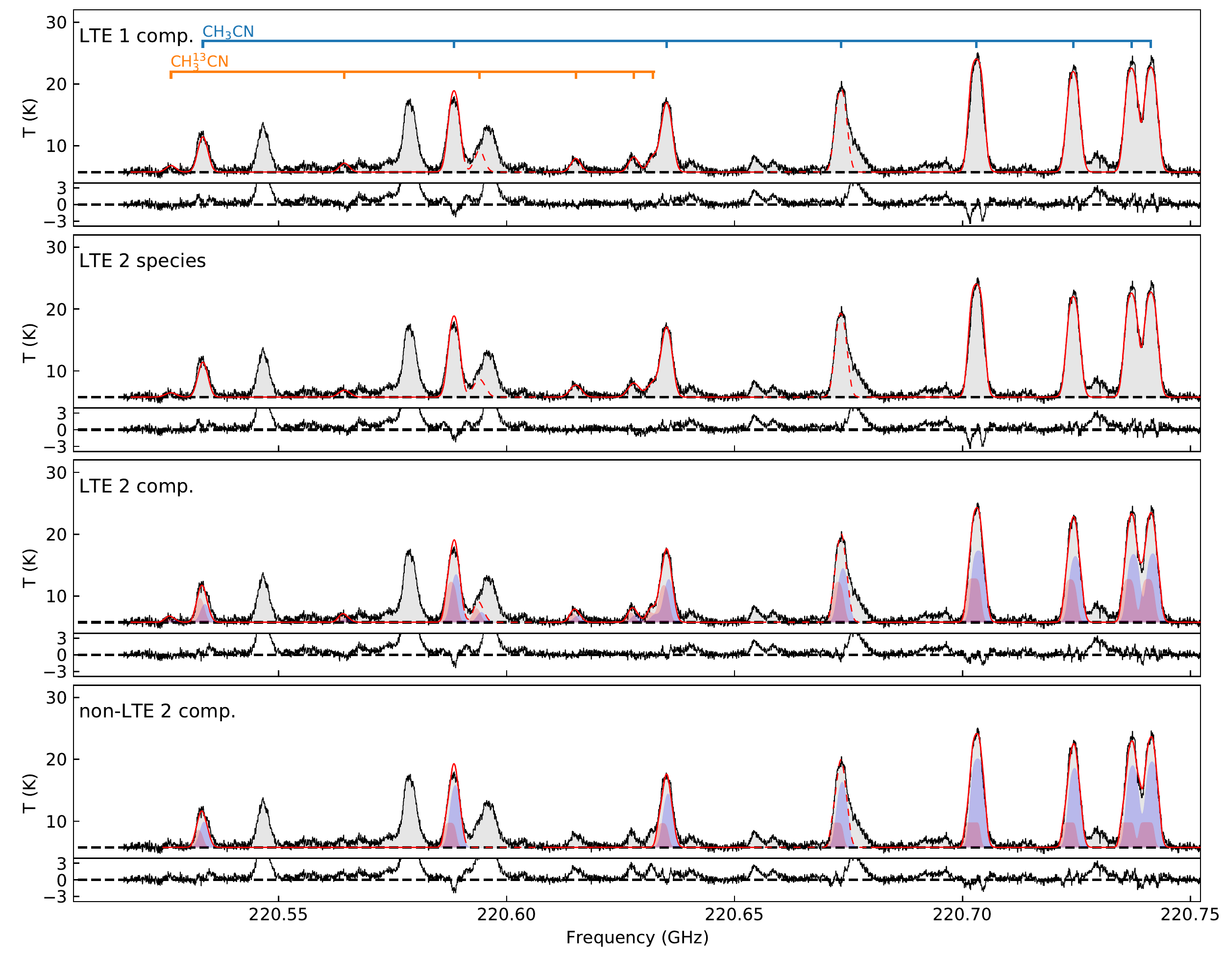}
\end{center}
\caption{Modeled \ce{CH3CN} and \ce{CH3{}^13CN} spectra overlaid on the observed spectra.
The first panel shows the LTE model one velocity component fitting. The second panel is the same as the first panel but treats \ce{CH3CN} and \ce{CH3{}^13CN} as two species with independent temperature, velocity and linewidth. 
The third panel is the LTE two velocity component fitting while the fourth panel shows the non-LTE one.
The red curve shows the modeled spectrum;
The K=4 component of \ce{CH3CN} and K=3 component of \ce{CH3{}^13CN} suffer from severe line contamination so they were not included in the fitting. The model in these frequency ranges is shown with the dashed line. 
The red and blue shaded areas in the two bottom panels represent the modeled spectrum from the individual components. The residual from the best-fit is shown in the bottom frame of each panel. The best-fit parameters are listed in Table \ref{tab:cassis} and the corner plots are shown in Appendix \ref{sec:mcmc}.
}
\label{fig:cassis}
\end{figure*}

\subsection{Kinematics of \ce{CH3CN}}
\label{sec:uv_gau}
For a spectrally resolved optically thin line, the peak position of the line emission is believed to represent the most likely location of the emitting gas \citep{sa87,ha13}.
To disentangle the velocity structure of the \ce{CH3CN}, we perform a uv-domain Gaussian fitting to the spatial emission at different velocity channels.
For \ce{CH3CN}, the thermal width of \ce{CH3CN} (12--11) at 200 K is $\sim$0.5 km s$^{-1}$.
Thus, we took a step of 0.5 km s$^{-1}$ to increase S/N while resolving the gas kinematics, and a velocity range from 6 to 10 km s$^{-1}$ (Figure \ref{fig:asym}). We use the K=3 and K=7 components which suffer less from contamination of other lines (Figure \ref{fig:spec}). 
The K=7 emission is most likely optically thin, and the
K=3 emission, although it is optically thick, has the highest S/N ratio among the K-ladder transitions.
Figure \ref{fig:uv_chan} (right) shows the fitted central positions from different velocities.
The emitting centers show similar trends in these two transitions;
the majority of the \ce{CH3CN} emission is associated with VLA4A which is considered to be the warmer source with a higher accretion rate \citep{hs19}. 
A clear velocity gradient from the west to the east is seen.
The direction of this velocity gradient is broadly consistent with the position angle of the circumstellar disk around VLA4A and the circumbinary disk \citep{di21}; in the literature, Ethylene Glycol (aGg'-\ce{(CH2OH)2}) is found to trace rotating gas from the circumstellar disk of VLA4A \citep{di21}.
By comparing these emitting centers to kinematic models, we find that the \ce{CH3CN} emission most likely associates with a disk or a rotating envelope around VLA4A rather than that from VLA4B or the circumbinary disk (see Appendix \ref{sec:rot}).
However, this analysis shows that the velocity gradient from \ce{CH3CN} cannot be interpreted by a simple rotationally supported disk or from angular momentum conservation.
It is then likely that \ce{CH3CN} and aGg'-\ce{(CH2OH)2} trace different gas components.
High-angular-resolution observations are required to confirm the origins of the \ce{CH3CN} emission.

\begin{table*}
    \centering
    \caption{\ce{CH3CN} fitting results\label{tab:cassis}}
    \begin{tabular}{ccccccccccccc}
    \hline\hline
Model   
& component
& $N_{\ce{CH3CN}}$
& T$_{\rm rot}$
& FWHM
& size
& log(n$_{H2}$)
& V$_{LSR}$	
& $\frac{\ce{CH3CN}}{\ce{CH3{}^13CN}}$ 
\\
& 		
& (10$^{16}$ cm$^{-2}$)		
& (K)
& (km s$^{-1}$)	
& (arcsec)
& (cm$^{-3}$)
& (km s$^{-1}$)	
& 	\\
\hline
LTE model & -  & 5.13$^{+0.07}_{-0.06}$ & 247.8$^{+1.4}_{-1.4}$ & 3.34$^{+0.01}_{-0.01}$ & 0.273$^{+0.001}_{-0.001}$ & - & 8.099$^{+0.003}_{-0.003}$ & 15.6$^{+0.2}_{-0.2}$\\
\hline
LTE two species & \ce{CH3CN}  & 5.05$^{+0.07}_{-0.07}$ & 246.6$^{+1.4}_{-1.4}$ & 3.33$^{+0.01}_{-0.01}$ & 0.274$^{+0.001}_{-0.001}$ & - & 8.094$^{+0.003}_{-0.003}$ & 25.2\tablefootmark{a}\\
& \ce{CH3{}^13CN}  & 0.20$^{+0.02}_{-0.01}$ & 166.8$^{+10.5}_{-10.9}$ & 3.91$^{+0.06}_{-0.06}$ & 0.27$^*$ & - & 8.117$^{+0.032}_{-0.032}$ & - \\
\hline
LTE 2-comp&
extend & 2.77$^{+0.08}_{-0.08}$ & 189.2$^{+2.9}_{-2.9}$ & 3.02$^{+0.02}_{-0.01}$ & 0.247$^{+0.002}_{-0.003}$ & - & 7.490$^{+0.023}_{-0.024}$ & 24.6$^{+1.0}_{-0.9}$\\
& compact & 11.0$^{+0.9}_{-0.8}$ & 275.8$^{+7.7}_{-8.0}$ & 2.67$^{+0.02}_{-0.02}$ & 0.154$^{+0.004}_{-0.003}$ & - & 8.950$^{+0.023}_{-0.024}$ & 16.84$^{+0.6}_{-0.6}$\\
\hline
non-LTE 2-comp&
extend & 1.74$^{+0.05}_{-0.04}$ & 211.9$^{+2.9}_{-2.9}$ & 3.00$^{+0.01}_{-0.01}$ & 0.287$^{+0.002}_{-0.003}$ & 
6.79$^{+0.04}_{-0.03}$ & 7.739$^{+0.014}_{-0.015}$ & - \\
& compact & 7.18$^{+0.44}_{-0.40}$ & 193.6$^{+13.6}_{-10.9}$ & 2.38$^{+0.04}_{-0.04}$ & 0.141$^{+0.004}_{-0.004}$ & 
$>$8.07$^{+0.40}_{-0.43}$\tablefootmark{b} & 9.321$^{+0.024}_{-0.027}$ & -\\ 
\hline
\citet{be20} & -  & 4 & 180 & 4.0\tablefootmark{*} & 0.3\tablefootmark{*} & - & - & -\\
\hline
\citet{ya21} & -  & 0.6$^{+0.18}_{-0.13}$ & 150\tablefootmark{c} & - & 0.5\tablefootmark{*} & - & - & -\\
\hline
    \end{tabular}
\tablefoot{
    \tablefoottext{*}{The parameter is fixed.}
    \tablefoottext{a}{The value is derived from the \ce{CH3CN}/\ce{CH3{}^13CN} column density ratio.}
    \tablefoottext{b}{The value is considered as a LTE case, because the posterior distribution reaches the upper boundary of $5\times10^8$ cm$^{-3}$ in the MCMC modeling given a critical density of $\sim10^8$ cm$^{-3}$ for \ce{CH3CN} J=12--11, K=0--7.}
    \tablefoottext{c}{The fitting is done with a step of 50 K.}
}
\end{table*}


\subsection{A streamer in DCN and \ce{C^{18}O} emission}
\label{sec:dcn}
\begin{figure}
\begin{center}
\includegraphics[width=0.5\textwidth]{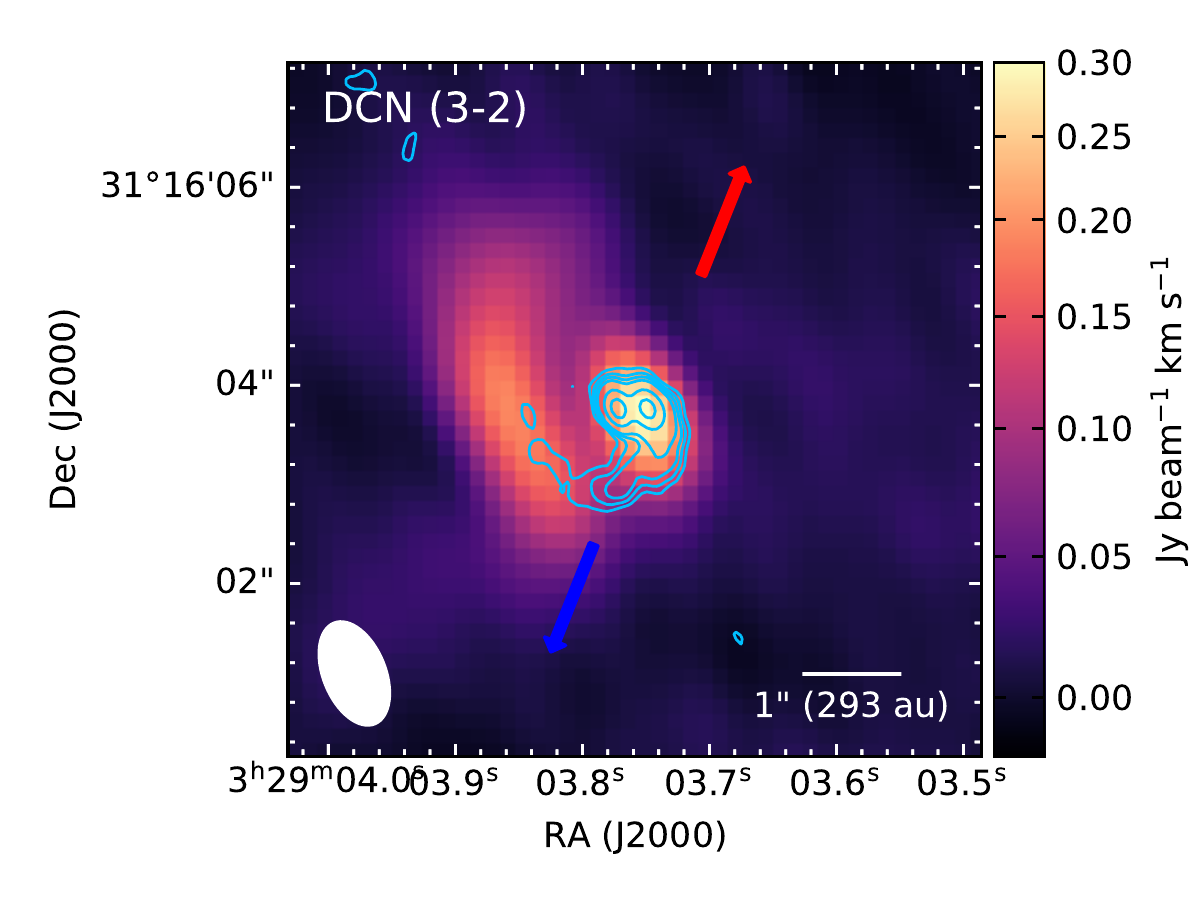}
\end{center}
\caption{Integrated intensity map of DCN (3-2) over a velocity range of 8.26-8.8 km s$^{-1}$. The contours represent the ALMA 1.3 mm continuum emission with levels at 3, 5. 7, 10, 30, 70$\sigma$. The blue and red arrows indicate the CO outflow directions from \citet{le17}.
}
\label{fig:dcn_mom0}
\end{figure}

Figure \ref{fig:dcn_mom0} shows the NOEMA DCN (3-2) integrated intensity map. The DCN emission traces the central binary system and reveals a streamer in the east.
Interestingly, this streamer matches the tail of the spiral structure that is seen in the continuum emission, and extends this spiral to outer regions at a few arcsec ($\sim2\farcs4$, 700 au).
This kind of streamers connecting inner and outer regions have been recently identified in embedded systems and are considered as infalling flows that provide the material supplying the growth of disks and central protostars \citep{pi20,al20,mu21,ca21,th22,va22}.

\begin{figure*}
\begin{center}
\includegraphics[width=0.96\textwidth]{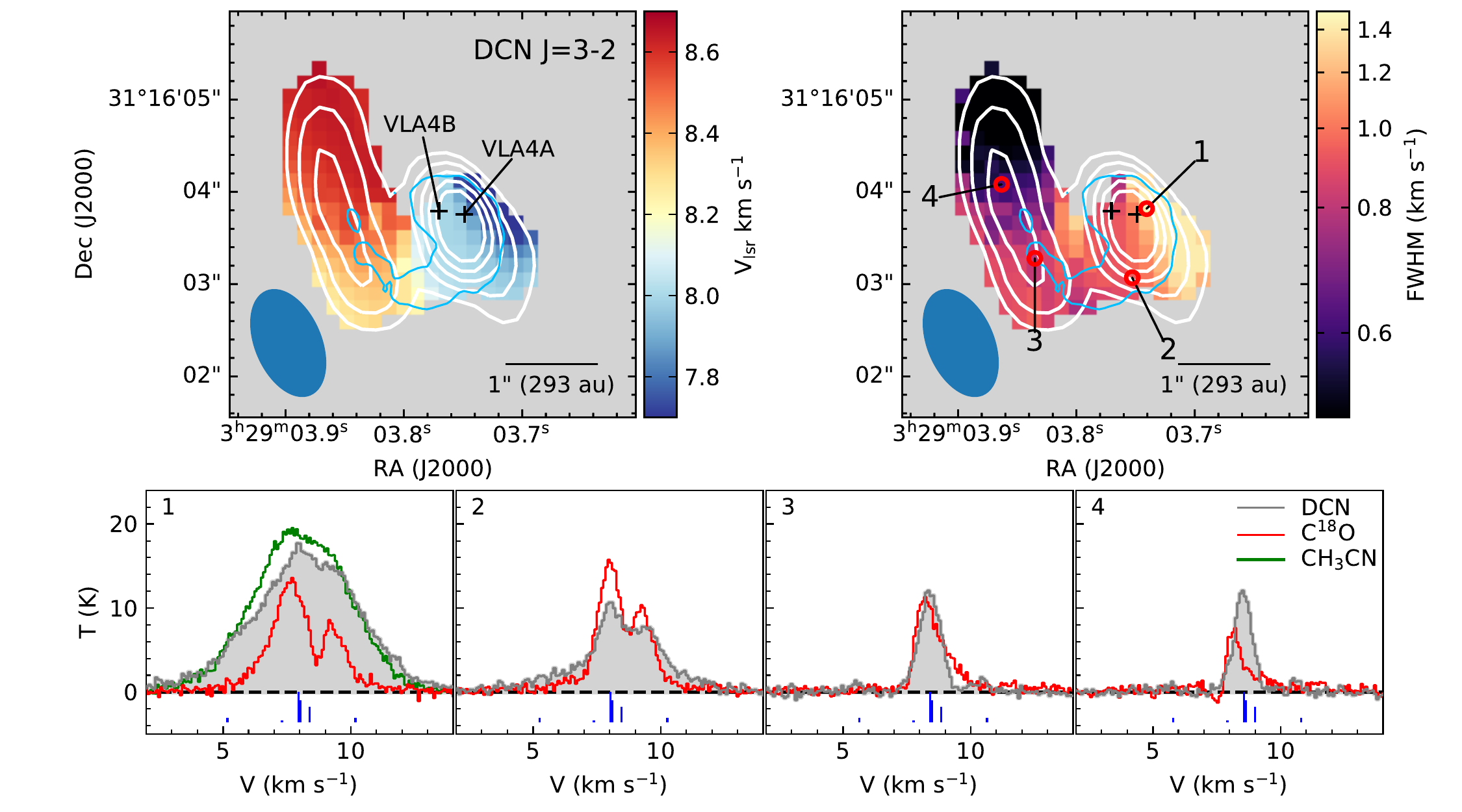}
\end{center}
\caption{
DCN (3-2) centroid-velocity ({\it top left}) and linewidth ({\it top right}) maps from the blue-shifted component in the hyperfine structure fitting (see Appendix \ref{sec:decom}). The white contours show its peak intensity. The contour levels start at 25$\sigma$ and increase in 5$\sigma$ step, where $\sigma=0.3$ K. The blue contour shows the ALMA 1.3 mm continuum emission at the 3$\sigma$ level. ({\it Bottom}) The spectra of DCN (3--2), \ce{C^{18}O} (2--1), and \ce{CH3CN} ($12_3--11_3$) toward the four positions marked in the right panel.
The blue bars show the relative intensities of the DCN (3--2) hyperfine structure at the velocity of the blue-shifted component (see Appendix \ref{sec:decom}).
}
\label{fig:dcn_fit}
\end{figure*}

Because the DCN spectral profile likely originates from a complex structure (Figure \ref{fig:dcn_fit}), we employ multiple component fitting in order to isolate the streamer and further study its kinematics. We fit the hyperfine structure of the DCN (3-2) line emission using PySpecKit \citep{gi11} with the line parameters from CDMS \citep{mu05,en16}, for which the DCN parameters are from \citet{br04} and references therein.
Although the satellite line emission is marginally detected (Figure \ref{fig:dcn_fit}), the weak emission in the streamer suggests that the DCN (3-2) emission is optically thin. 
Toward the center, the hyperfine structure is not clearly seen with the complex and broad line emission. 
However, analyzing this DCN structure toward the center is beyond the scope of this paper. 
Thus,
we assume a total optical depth of 0.1 in the hyperfine structure fitting (HFS).
Toward the center region, several velocity components are revealed from the spectra. We therefore perform hyperfine fitting with  multiple-velocity components (up to three, see Appendix \ref{sec:decom}).
We only focus on the regions where the peak intensity has a S/N$>$25 from the spectrum to distinguish the streamer from the more extended emission background.
Figure \ref{fig:dcn_fit} shows the maps of the fitted centroid velocity and linewidth of the blue-shifted component which traces the streamer.
A small velocity gradient $\sim$0.12 km s$^{-1}$ arcsec$^{-1}$ along the streamer with a relatively narrow linewidth of $\sim0.7$ km s$^{-1}$ is revealed.
We note that a tiny velocity break from 8.4 km s$^{-1}$ to 8.1 km s$^{-1}$ comes from a change of fitting with 2-velocity to 3 velocity components (see Appendix \ref{sec:decom}).

\begin{figure*}
\begin{center}
\includegraphics[width=0.96\textwidth]{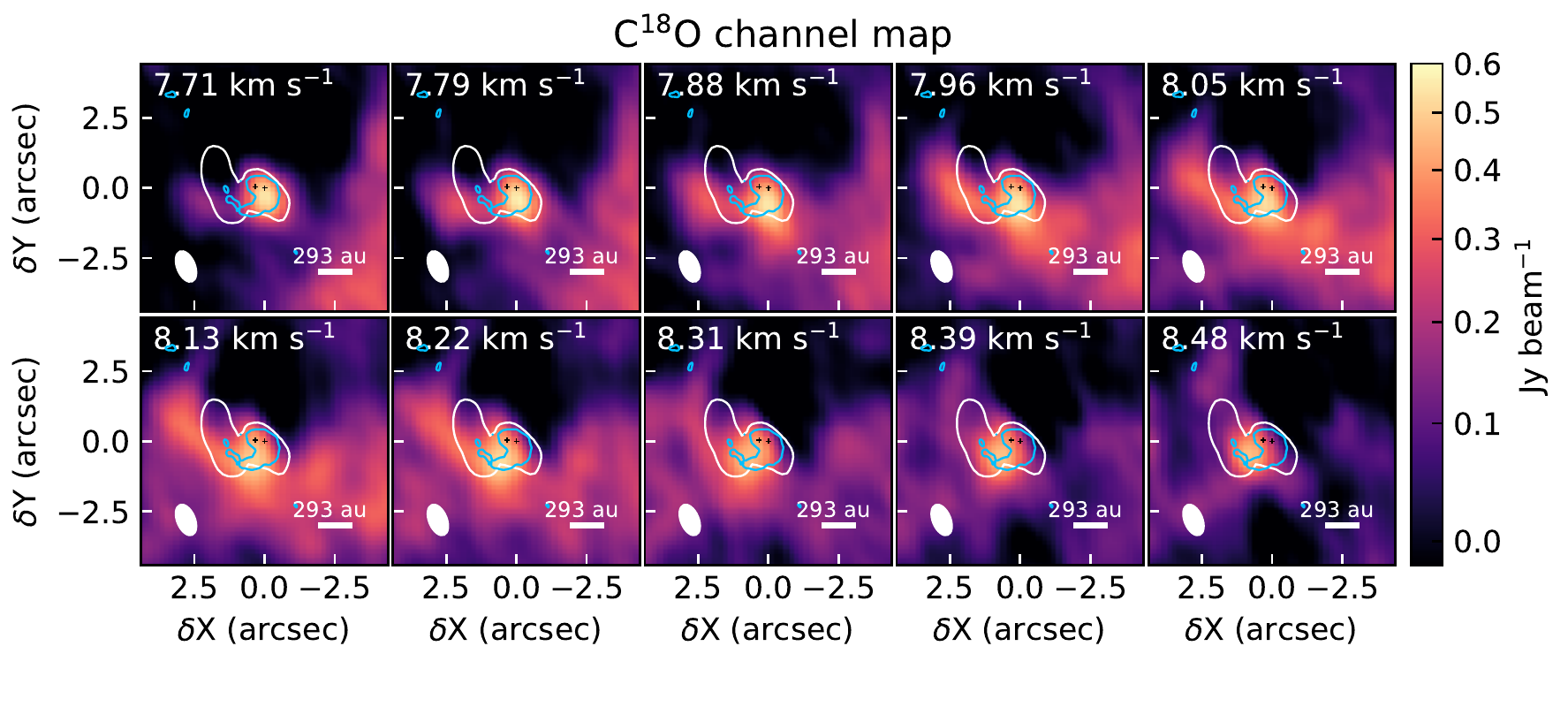}
\end{center}
\caption{\ce{C^{18}O} (2-1) channel map with the velocity range from 7.7-8.5 km s$^{-1}$. The blue contour shows the ALMA continuum emission at 3$\sigma$ and the white contour shows the DCN (3-2) peak intensity map at 25$\sigma$}
\label{fig:c18o_chan}
\end{figure*}

Using ALMA, \citet{to18} find that the dusty spiral from the continuum emission well matches the \ce{C^{18}O} (2--1) blue-shifted emission (6.0--8.5 km s$^{-1}$, in their Figure 15).
Figure \ref{fig:c18o_chan} shows the NOEMA \ce{C^{18}O} (2--1) channel maps at 7.7--8.5 km s$^{-1}$ in comparison with the streamer traced by the DCN and continuum emission. Although at the velocity of the DCN streamer ($\sim8.3-8.6$ km s$^{-1}$, Figure \ref{fig:dcn_fit}) the \ce{C^{18}O} emission is offset from the \ce{DCN} emission, the emission peak matches well the inner spiral at a velocity $\leq8.2$ km s$^{-1}$; and this velocity well connects the DCN blue-shifted end from $\sim$8.3 km s$^{-1}$ to a velocity of $\sim7.7$ km s$^{-1}$ in the inner region (see also Figure \ref{fig:alma_c18o} from the ALMA data).
The channel maps of DCN/\ce{C^18O} in the region of interest are provided in Appendix \ref{sec:dcn_chan}.

Figure \ref{fig:dcn_fit} also shows the DCN and \ce{C^{18}O} line profiles at four selected positions from the streamer.
In the outer part of the streamer (i.e., position 4), the \ce{C^{18}O} emission is offset from the DCN emission in velocity.
In the middle part of the streamer (i.e., position 3), the DCN and \ce{C^{18}O} lines have a similar centroid velocity.
At position 2, while the DCN emission becomes relatively weak compared with that of \ce{C^{18}O}, the lines show similar profiles.
It seems that, along the streamer, the physical properties change, producing a different DCN/\ce{C^{18}O} integrated intensity ratio.
However, the velocity gradient revealed by DCN emission is broadly consistent with that of \ce{C^{18}O} at velocities $\lesssim$8.3 km s$^{-1}$ (Figure \ref{fig:alma_c18o}), and might trace infall of material.
Higher spatial resolutions observations are needed to test for infall via modeling.

\begin{figure}
\begin{center}
\includegraphics[width=0.5\textwidth]{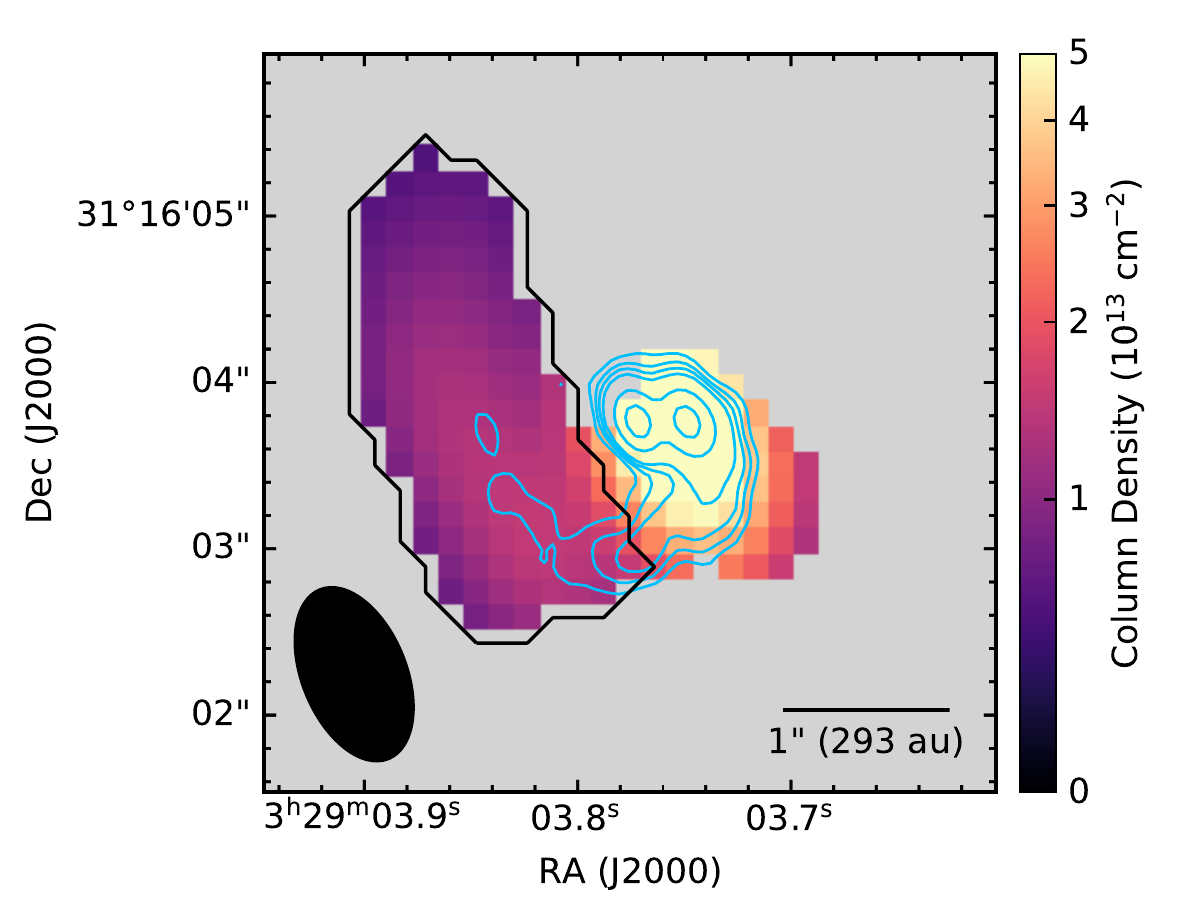}
\end{center}
\caption{DCN column density map. The blue contours represent the ALMA 1.3 mm dust continuum emission. The black polygon encloses the area that is used to measure the mass of the streamer.
}
\label{fig:CD}
\end{figure}

To estimate the mass of the streamer, we calculate the DCN column density (Figure \ref{fig:CD}) over a velocity range of 7--10 km s$^{-1}$ targeting the streamer. The column density is calculated under assumptions of LTE and optically thin emission (i.e., eq.\ 80 in \citealt{ma15}).
Due to the lack of measurement for the DCN collision rates, 
we consider those for HCN \citep{va20}. The critical density of HCN (3--2) is $1.0\times10^7$ cm$^{-3}$ at 20 K and $5.7\times10^6$ cm$^{-3}$ at 50 K in the optically thin limit \citep{sh15}. This is close to the lower bound estimated from previous works n$_{H_2}\geq10^6$ cm$^{-3}$ within 600 au using \ce{H2CO} \citep{bi17}. Thus, the LTE assumption might cause an underestimate of the actual streamer mass.
The excitation temperature is estimated considering the luminosity of SVS13A (45.3 L$_\odot$) and the projected distance to VLA4A 
(i.e., $T_{\rm d}=\frac{2.4\times10}{(r/pc)^{2/5}}(\frac{L}{10^6 L_\odot})^{1/5}$,
\citealt{go74}). This gives a temperature of $\sim35-45$ K in the streamer.
To exclude the extended DCN emission, we employ a mask with a peak emission above 25 $\sigma$ and a linewidth $<$ 1 km s$^{-1}$ from the hyperfine fitting (Figure \ref{fig:dcn_fit}). 
This gives a conservative lower limit to the streamer area.
Figure \ref{fig:CD} shows the DCN column density of the streamer. Adopting the same LTE excitation state  for \ce{C^18O}, we further estimate the abundance ratio of [DCN]/[\ce{C^{18}O}] to be $\sim1.36\times10^{-3}$ at position 3 in Figure \ref{fig:dcn_fit}. We note that with a different $T_{\rm ex}$=20--80 K, this abundance ratio varies from $(1.54-1.28)\times10^{-3}$.
Taking the canonical ISM value of CO/\ce{C^{18}O}$\sim$560 and CO abundance of 10$^{-4}$ \citep{wi94}, we find a DCN abundance of $\sim2.4\times10^{-10}$ with respect to H$_2$.
As a result, the streamer mass is estimated to be $\gtrsim$0.0042 M$_\odot$.

\section{Discussion}
\label{sec:dis}
\subsection{Physical and chemical conditions from \ce{CH3CN}}
Table \ref{tab:cassis} lists the physical conditions of SVS13A derived from the \ce{CH3CN} K-ladder with different models.
Besides the main \ce{CH3CN} isotopologue, the high-sensitivity data also allowed us to detect the isotopologue \ce{CH3{}^13CN} (Figure \ref{fig:spec}).
With the LTE 1-component model, the abundance ratio \ce{CH3CN}/\ce{CH3{}^13CN} is found to be $\sim20$. This is surprisingly low given the interstellar isotopologue ratio [$^{12}$C]/[$^{13}$C]$\sim$68 \citep{mi05}.
In IRAS16293, \citet{ca18} find the column density ratios \ce{CH3CN}/\ce{CH3{}^13CN}$\sim$67 and \ce{CH3CN}/\ce{^{13}CH3CN}$\sim$77. 
\ce{^{12}C}/\ce{^{13}C} ratio can be affected by selective photodissociation and fractionation.
For example, the \ce{^12CO}/\ce{^13CO} ratio is found to vary by a factor of $>$4 from the inner to outer parts of the TW Hya disk \citep{yo22}.
As selective photodissociation is suggested to affect the isotopologue ratios in the disk atmosphere \citep{mi14,sm20}, this mechanism can decrease \ce{^{12}C}/\ce{^{13}C} ratio by destruction of the main isotopologue.
On the other hand, isotopic fractionation reaction can enhance the \ce{^13C} abundances in molecules but usually happen in cold region with exothermic reactions \citep{ro15}. 
It is noteworthy that, in SVS13A, \ce{CH3CN} is suggested to be synthesised via gas-phase reactions during cold prestellar phase \citep{bi22a}.
Alternatively, the \ce{CH3CN} column density is underestimated due to an inaccurate opacity correction.
This scenario is similar to that reported by \citet{di21} with the \ce{^{13}CH3OH}/\ce{CH3{}^{18}OH}$\sim$3, significantly smaller than the expected value $\sim$7.
Broadly speaking, the optically thick \ce{CH3CN} emission (which depends on K-transition) and the optically thin \ce{CH3{}^13CN} emission can trace a different region/layer.

Thus, we conducted a fitting procedure 
assuming that \ce{CH3CN} and \ce{CH3{}^13CN} are independent species (i.e., different $V_{\rm lsr}$, $\Delta$V, $T_{\rm ex}$ and $N_{\rm mol}$, Table \ref{tab:cassis} LTE two species).
However, the source size of \ce{CH3{}^13CN} was fixed to $0\farcs27$ to match that of \ce{CH3CN}.
Given that the emission of \ce{CH3{}^13CN} is optically thin, the derived column density is inversely proportional to the square of the source size.
As a result, the temperature of \ce{CH3{}^13CN} is significantly lower than that of \ce{CH3CN}.
This result (Table \ref{tab:cassis}) implies that the \ce{CH3{}^13CN} with a smaller optical depth traces a region different from \ce{CH3CN}, which might explain the small column density ratio.

We also employ a two velocity-component model, in an attempt to resolve the issues of optical depth (K=0-3) and asymmetric line profile (Figure \ref{fig:asym}).
A relatively optically thin blue-shifted component ($\tau=0.28-3.2$) and an optically thick ($\tau=0.8-6.4$) redshifted component are found (Figure \ref{fig:cassis}).
The residual from the optically thick K=3 component is now significantly decreased.
Table \ref{tab:cassis} lists the fitted parameters from the MCMC fitting as shown in Figure \ref{fig:cassis}.
The result implies that an optically thick compact component was missed in the one-component model while the optically thin component can dominate the line emission.
Because of the missing of the optically thick component due to insufficient spatial resolution, the column density of the \ce{CH3CN} can be severely underestimated.
It is noteworthy that, with the two-velocity component model, the isotopologue ratios of $\frac{\ce{CH3CN}}{\ce{CH3{}^13CN}}$
($\sim$25 and $\sim$17)
are still too low compared to the canonical value ($\sim$68; \citealt{mi05}).
Therefore, the fitting procedure taking \ce{CH3{}^13CN} as an independent species is still required.

Previous works conducted non-LTE analysis using different chemical tracers to estimate the H$_2$ density ($n_{\rm H_2}$) toward SVS13A.
Using the IRAM 30m telescope, \citet{co16} estimate a H$_2$ volume density of $\geq3\times10^7$ cm$^{-3}$ inside a small radius of 25 au with a temperature of 150-260 K through HDO observations. This compact source is believed to originate from a hot corino. With \ce{CH3OH}, \citet{bi17} found a similarly compact component with $n_{\rm H_2}>10^8$ cm$^{-3}$ but lower $T_{\rm kin}\sim80$ K within a radius $\sim$35 au, and an additional extended component with $n_{\rm H_2}>10^6$ cm$^{-3}$ and $T_{\rm kin}<70$ K within $\sim$350 au. Recently, \citet{co21} suggest that the \ce{SO}, \ce{SO2}, and \ce{H2CS} emission from SVS-13A is confined within the hot corino, with a size 60--120 au, $n_{\rm H_2}\geq1\times10^5-6\times10^6$ cm$^{-3}$ and $T_{\rm kin}=100-300$ K. 
In our non-LTE two-component models of \ce{CH3CN}, we find a compact component with a size of $0\farcs14$ (40 au) with $n_{\rm H_2}>10^8$ cm$^{-3}$ (LTE) and an extended component with a size of $0\farcs29$ (85 au) and $n_{\rm H_2}\sim6\times10^6$ cm$^{-3}$.
Our result is broadly consistent with the literature;
most likely, SVS13A consists of a compact high-density hot corino with a diameter of $50-70$ au, and an extended warm structure up to $\sim120$ au.
It is noteworthy that with the high-resolution ALMA observation, \citet{di21} decompose SVS13A into three disks, VLA4A ($67\times62$ au), VLA4B (unclear from line observation), and a circumbinary disk ($240\times180$ au).

\citet{co16} report the HDO observation toward SVS13A; the non-LTE analysis gives the best-fit column density of $4\times10^{17}$ cm$^{-2}$, and a temperature of 200 K with a source size of $0\farcs2$. The source size as well as the linewidth of $\sim4.5$ km s$^{-1}$ (\ce{H2O}, \citealt{kr12}) are broadly consistent with that of \ce{CH3CN}.
Further, \citet{co16} found an upper limit of water column density $\leq8\times10^{17}$ cm$^{-2}$ based on a \ce{H2{}^{18}O} non-detection.
If we assume that the \ce{CH3CN} compact component originates from the same region as the HDO and water,
we obtain lower limits of \ce{CH3CN} abundance relative to \ce{H2O} $\geq$18-26\% for the compact component (Table \ref{tab:cassis}).
This is higher than the values reported in comets (Hyakutake: 0.01\%; Hale-Bopp: 0.02\%; 2001 A2: 0.028\%; 73P/SW3: 0.030\%; 67P: 0.59\%) and between interstellar sources (IRAS 16253: 0.25\%; Sgr B2(N): 30\%) (\citealt{mu11,bo17,ru19}).

\subsection{Accretion flow traced by \ce{DCN}?}
\label{sec:acc}
During the embedded phase of a protostar, infall of material plays a crucial role in supplying material to the central source and protoplanetary disk. 
Recent studies have detected large infalling streamers at an early stage, which are believed to affect the protostellar accretion \citep{pi20,al20,ca21,gi21,mu21,ga22,th22,va22}.

In SVS13A, the streamer is seen in DCN, \ce{C^{18}O} and partially in dust continuum emission (see section \ref{sec:dcn}), connecting the outer envelope with the central binary.
At early evolutionary stages, large infalling streamers (1000s of au) are identified via Carbon-chain species such as \ce{HC3N} toward Per-emb-2 \citep{pi20} and \ce{HC5N} toward IRAS16293-2422 \citep{mu21}; these molecules usually trace less chemically evolved regions, while the 
deuterated species DCN traces relatively chemically evolved materials.
The fresh materials, together with the enormous streamers, in the literature suggest that these streamers originate from outside the cloud core.
In contrast, the streamer toward SVS13A via DCN is much smaller $\sim700$ au. It is unclear if this streamer is extended farther away or not due to the insufficient sensitivity and/or lack of proper tracers.

The velocity gradient along the streamer revealed by DCN and \ce{C^{18}O} suggests it is related to infalling material towards the protostars. The total mass of this streamer is estimated to be $M_{\rm streamer}\sim0.0042 M_\odot$ (section \ref{sec:dcn} and Figure \ref{fig:CD}). Assuming a free-fall collapse, the mass infalling rate from the streamer can be estimated following \citet{pi20}:
\begin{equation}
\dot{M}_{\rm infall}=\frac{M_{\rm streamer}}{t_{ff}},
\end{equation}
where $t_{ff}$ is the free-fall timescale.
Here $t_{ff}$ can be obtained given the enclosed mass $M$ within a radius $R$, i.e., 
\begin{equation}
t_{ff}=\sqrt{\frac{R^3}{GM}}.
\end{equation}
By measuring the proper motion, \citet{di21} derive a total mass of VLA4A/B as 1 M$_\odot$ from the orbital motion.
Given the detected farthest projected distance of the DCN streamer $\sim$700 au, we derive a free-fall time of $\sim$3000 yr. We note that this can be taken as an upper limit since the initial velocity of DCN is ignored in the calculation \citep{va22}.
As a result, we find a mass infalling rate of $\geq1.4\times10^{-6}$ M$_\odot$ yr$^{-1}$ if this streamer is in a free-fall motion.
This value is similar to that of the streamers found in other Class I protostars (Per-emb-2: 10$^{-6}$ $M_\odot$ yr$^{-1}$ \citealt{pi20}; Per-emb-50: 1.3$\times$10$^{-6}$ $M_\odot$ yr$^{-1}$ \citealt{va22} ).
The infalling streamer appears to connect to the bursting source VLA4A from the ALMA high-resolution continuum image \citep{di21}, for which the circumstellar disk has a mass of 0.004-0.009 $M_\odot$.
It would take less than 2800--6400 years for the infalling streamer to replenish this disk. 
If the streamer supplies the material to the circumbinary disk with $M_{\rm disk}=0.052$ $M_\odot$, a replenishment could happen in a timescale of $<$37000 yr. These timescales are broadly consistent with the recurrence timescale of $10^3-10^5$ yr for accretion bursts \citep{sc13,co19,fi19,hs18,hs19,pa21}, for which the timescale is believed to be smaller in earlier stages.

The high luminosity ($L\sim45.3~L_\odot$) of SVS13A suggests that it is undergoing an accretion burst with a protostellar mass accretion rate of $2.8\times10^{-5}$ M$_\odot$ yr$^{-1}$ \citep{hs19}.
As a result, the infalling rate from the envelope to the disk contributed via the streamer can be $\geq5\%$ of the protostellar mass accretion rate.
If we assume the streamer dominates the mass infalling rate and the disk mass is constant, this implies that the protostar spends $>5$\% of the time in the burst phase, i.e., ``duty cycle'' in episodic accretion \citep{au14}.
However, if a significant mass is accreted onto the central source during the quiescent phase, this will reduce this value.
For comparison, by modeling the luminosity distribution, \citet{ev09} derive a duty cycle of 7\% and \citet{du12} estimate a value of 1.3\%.
We note that SVS13A appears to be a binary system (or perhaps triple, \citealt{le17}) and it is still unclear if the infalling streamer is feeding directly the disk around VLA4A or the circumbinary disk without higher-resolution observations.

\subsection{The structure of SVS13A}
\label{sec:not GIdisk}
The origin of the COMs in star forming regions is not fully understood.
COMs are typically found to trace inner compact regions where the temperature is higher than $\sim100$ K. These regions are called hot corinos in low-mass star forming regions \citep{ce04}. 
However, COMs can also be present in the shocked gas near the centrifugal barrier \citep{cs18} and irradiated cavity walls of outflows \citep{dr15}.
SVS13A contains a central binary (VLA4A and VLA4B), circumbinary disk, and a few spiral features (Figure \ref{fig:cont}, also see \citet{di21}).
Furthermore, the presence of infalling streamers is suggested to produce shocked gas, enriching the chemistry \citep{bi22b} and causing the increase of velocity dispersion \citep{di21} in VLA4A.
Toward SVS13A, the close binary with the spiral structure seen in continuum emission was considered as a fragmenting disk due to gravitational instability \citep{to18}.
However, through DCN emission, we find that the spiral is connected to an infalling streamer that extends to an outer region ($\sim700$ au) of the protostellar envelope.
Although the size of the \ce{CH3CN} emitting region ($0\farcs27$, Table \ref{tab:cassis}) is comparable to the size of VLA4A's circumstellar disk ($\sim0\farcs23$ in diameter, \citealt{di21}), the \ce{CH3CN} line profile (position 1 in Figure \ref{fig:dcn_fit}) suggests that the emission does not only come from a single component.
The two-component model further supports this hypothesis; the \ce{CH3CN} emission, at different velocities, originates from regions with different column density and temperature (or \ce{H2} density).
In the two-component model, the blue-shifted component dominates the line emission and has a velocity relatively close to the rotating envelope/disk surrounding VLA4A ($V_{\rm lsr}=7.36$ km s$^{-1}$) seen via aGg'-\ce{(CH2OH)2} by \citet{di21}. This also corresponds to the velocity
of the DCN and \ce{C^{18}O} streamer at the landing site, suggesting that the accretion from the large scale affects the circumstellar material. 
On the other hand, the optically thick component at the red-shifted side is not present in the aGg'-\ce{(CH2OH)2} line. A possibility is that it comes from the presumed primary VLA4B ($V_{\rm lsr}=9.33$ km s$^{-1}$). 
Kinematically, the central positions of \ce{CH3CN} K=3 and K=7 emission at different velocities also reveal a velocity structure that cannot be explained by a simple rotationally supported disk (Figures \ref{fig:uv_chan} and \ref{fig:rot_curve}).
Therefore, we suggest that \ce{CH3CN} as a COM not only traces disk material, but is also affected by the large-scale infall of material provided by the streamer.
This scenario is also proposed by \citet{bi22b} based on the elongated COM emission toward both VLA4A/4B.
Supporting this scenario, in the case of the close Class 0 binary IRAS 16293-2422 A, the COM emission was resolved down to a scale of 10 au, and displayed several peaks  outside the individual compact disks \citep{ma20}. These peaks, located at about 30 au from the protostars, presumably trace shocks related to the accretion of the circumbinary material into the individual Keplerian disks \citep{mo19,ma20}.

We find the large-scale infalling streamer funneling material to the protostellar system.
The infalling streamer seen via DCN, \ce{C^{18}O} and continuum emission lands closer to VLA4A (both spatially and spectrally), the source undergoing an accretion burst.
For example, toward the binary system [BHB2007] 11A, \citet{al19} find a filamentary structure 
that connects the central binary system to the circumbinary disk.
However, at the current spatial resolution, it is unclear if the infalling streamer in SVS13A lands on the disk surrounding VLA4A or the circumbinary disk. A higher resolution observation is required to disentangle the complex structure in the central region.

\section{Conclusion}
\label{sec:con}
We present the \ce{CH3CN} ($12_{\rm K}-11_{\rm K}$), \ce{CH3{}^13CN} ($12_{\rm K}-11_{\rm K}$), DCN (3-2), and \ce{C^{18}O} (2-1) observations toward SVS13A (Per-emb-44) from the NOEMA PRODIGE Large Program.
Through the line emission, we study the dynamics and physical conditions of the system. Our main conclusions are summarized as follows:
\begin{enumerate}

\item Through MCMC modeling, we find a low \ce{CH3CN}/\ce{CH3{}^13CN} ratio of 16 much lower than the isotope ratio value. We suggest a simple one-velocity component assumption is affected severely by optical depth, especially in a complex structure as SVS13A. Properly measuring the column density requires data including multiple transitions with different optical depths and high-spatial/spectral resolution to decompose the line emission.

\item The \ce{CH3CN} line profile 
is better explained by a two-component velocity model.
By splitting it into two components, we find \ce{CH3CN} can trace regions with dramatic difference in density and/or temperature.
The kinematics show that the \ce{CH3CN} emission can be affected by the infalling material from a streamer connecting the protostellar system to the outer envelope.

\item 
Contrary to the previous interpretation that the spiral seen via the continuum emission is part of a gravitationally unstable disk, we find that it is in fact connected to a much larger feature.
A streamer, possibly infalling, with a length of $\sim$700 au is identified;
it is partially seen through multiple tracers including DCN, \ce{C^{18}O}, and continuum emission. 
Under the assumption that the streamer is infalling,
it contributes a mass infalling rate onto the protostellar system of $\geq1.4\times10^{-6}$ M$_\odot$ yr$^{-1}$, $\geq5\%$ of the current protostellar mass accretion rate, $2.8\times10^{-5}$ M$_\odot$ yr$^{-1}$.
VLA4A might spend $\sim5\%$ of time in the bursting phase of episodic accretion.

\end{enumerate}

Our results show that the \ce{CH3CN} and \ce{CH3{}^13CN} emission traces hot gas in a complex structure in SVS13A. This region is connected to a streamer with a length of $\sim$700 au. If this streamer is feeding the central protobinary system, it can contribute a significant mass accretion rate. We speculate that the complexity might be caused by or associated to the large scale streamer. Observations with a higher spatial resolution are required to disentangle it.

\begin{acknowledgements}
We are grateful for the anonymous referee for the thorough and insightful comments that helped to improve this paper significantly.
The authors thank Dr. Emmanuel Caux, Dr. Valerio Lattanzi, Dr. Silvia Spezzano, and Dr. Christian Endres for valuable discussion in the line analyzing using CASSIS and CDMS.
T.-H. H., D. S.-C., J.E.P., P.C., M.T.V, and M.J.M.
acknowledge the support by the Max Planck Society. D. S.-C. is supported by an NSF Astronomy and Astrophysics Postdoctoral Fellowship under award AST-2102405.
A.F. acknowledges Spanish MICIN for funding support from PID2019-106235GB-I00.
NC acknowledges funding from the European Research Council (ERC) via the ERC Synergy Grant ECOGAL (grant 855130), and from the French Agence Nationale de la Recherche (ANR) through the project COSMHIC (ANR-20- CE31-0009).
MT acknowledges partial support from project PID2019-108765GB-I00 funded by MCIN/AEI/10.13039/501100011033.
This work is based on observations carried out under project number L19MB with the IRAM NOEMA Interferometer. IRAM is supported by INSU/CNRS (France), MPG (Germany) and IGN (Spain).
This paper makes use of the following ALMA data: ADS/JAO.ALMA\#2013.1.00031.S. ALMA is a partnership of ESO (representing its member states), NSF (USA) and NINS (Japan), together with NRC (Canada), MOST and ASIAA (Taiwan), and KASI (Republic of Korea), in cooperation with the Republic of Chile. The Joint ALMA Observatory is operated by ESO, AUI/NRAO and NAOJ.
\end{acknowledgements}

\bibliographystyle{aa}


\appendix
\setcounter{figure}{0}
\setcounter{table}{0}
\counterwithin{figure}{section}
\counterwithin{equation}{section}
\counterwithin{table}{section}

\section{MCMC fitting to \ce{CH3CN}/\ce{CH3{}^13CN} spectrum}
\label{sec:mcmc}
\begin{figure*}
\begin{minipage}[c]{0.48\textwidth}
\includegraphics[width=1\textwidth]{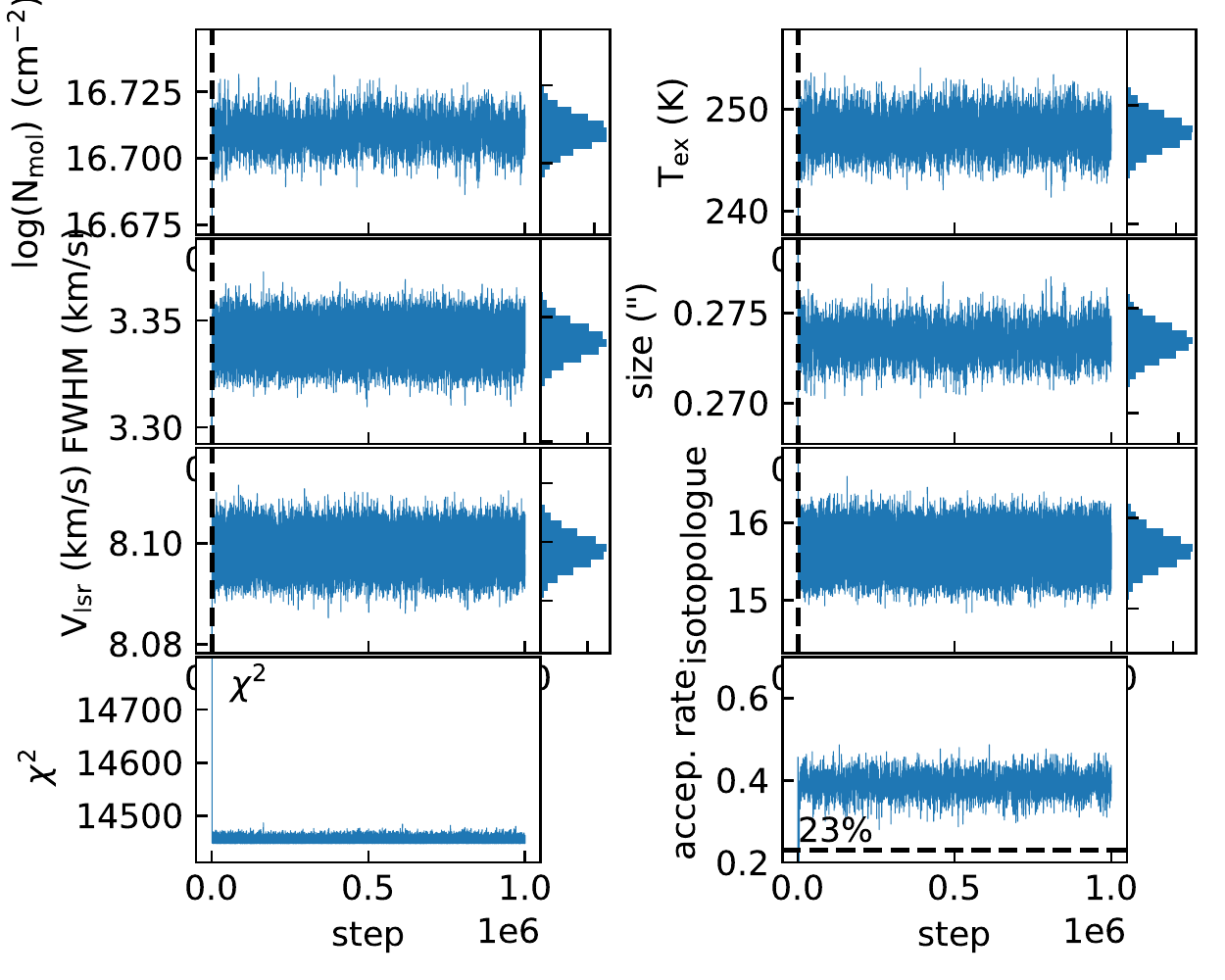}
\caption{Trace plot of the MCMC fitting for the LTE 1-component model. Each panel is for one physical parameter, for which the panel ``isotopologue'' is the column density ratio $\frac{\ce{CH3CN}}{\ce{CH3{}^13CN}}$. The vertical dashed line represents the cut before which the data points are discarded and the histogram in the right is plotted based on the saved sample. The bottom panels show the $\chi^2$ derived by CASSIS and acceptance rate measured in every 300 steps.}
\label{fig:trace}
\end{minipage}\hfill
\begin{minipage}[c]{0.48\textwidth}
\includegraphics[width=1\textwidth]{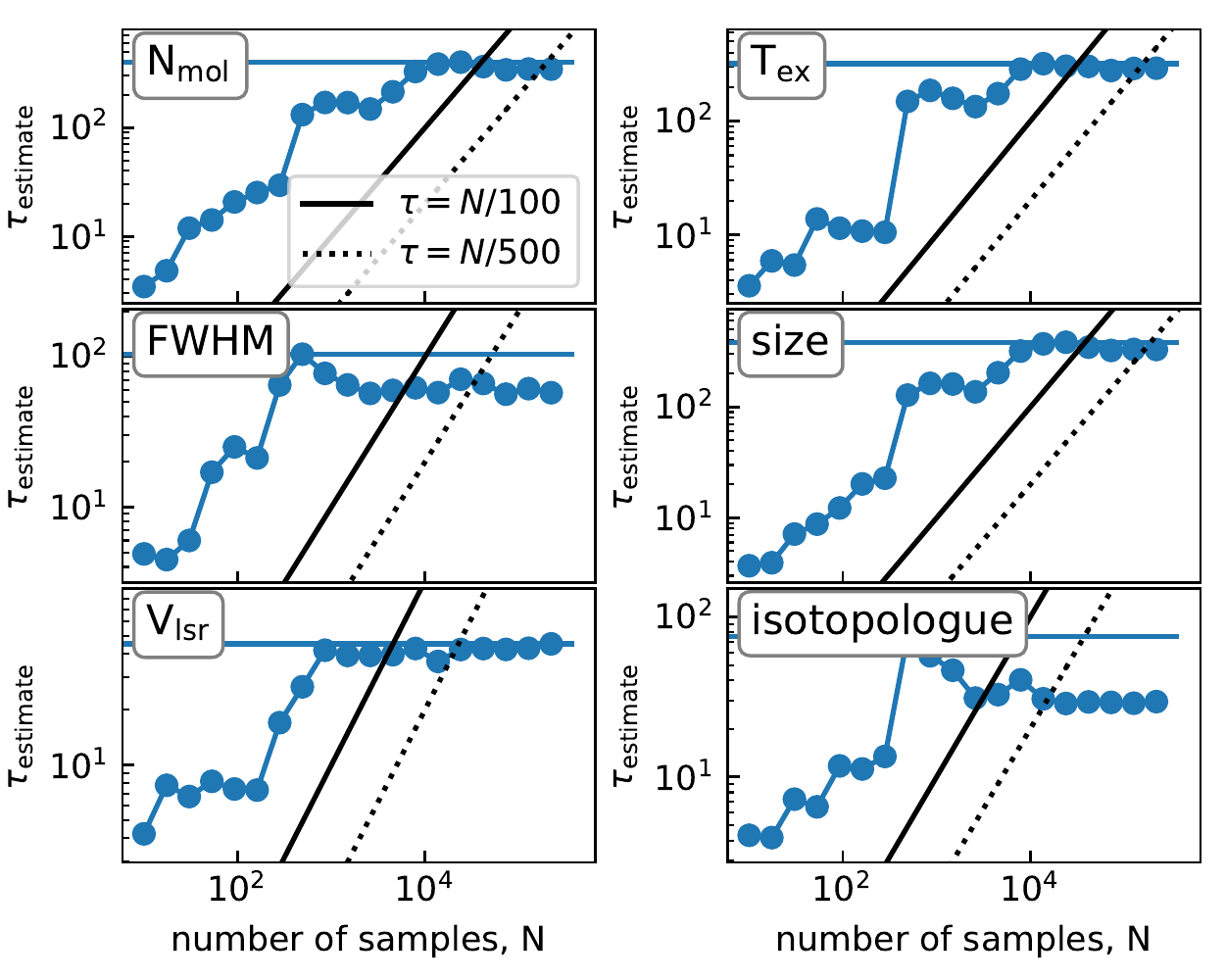}
\caption{Autocorrelation time as a function of number of samples for estimation for the LTE 1-component model in the MCMC fitting. The horizontal line represents the maximal autocorrelation time ($\tau$), and the solid and dashed lines represent the number of sample with 50 and 500 times of $\tau$.}
\label{fig:acf}
\end{minipage}
\end{figure*}

\begin{figure*}
\begin{center}
\includegraphics[width=0.97\textwidth]{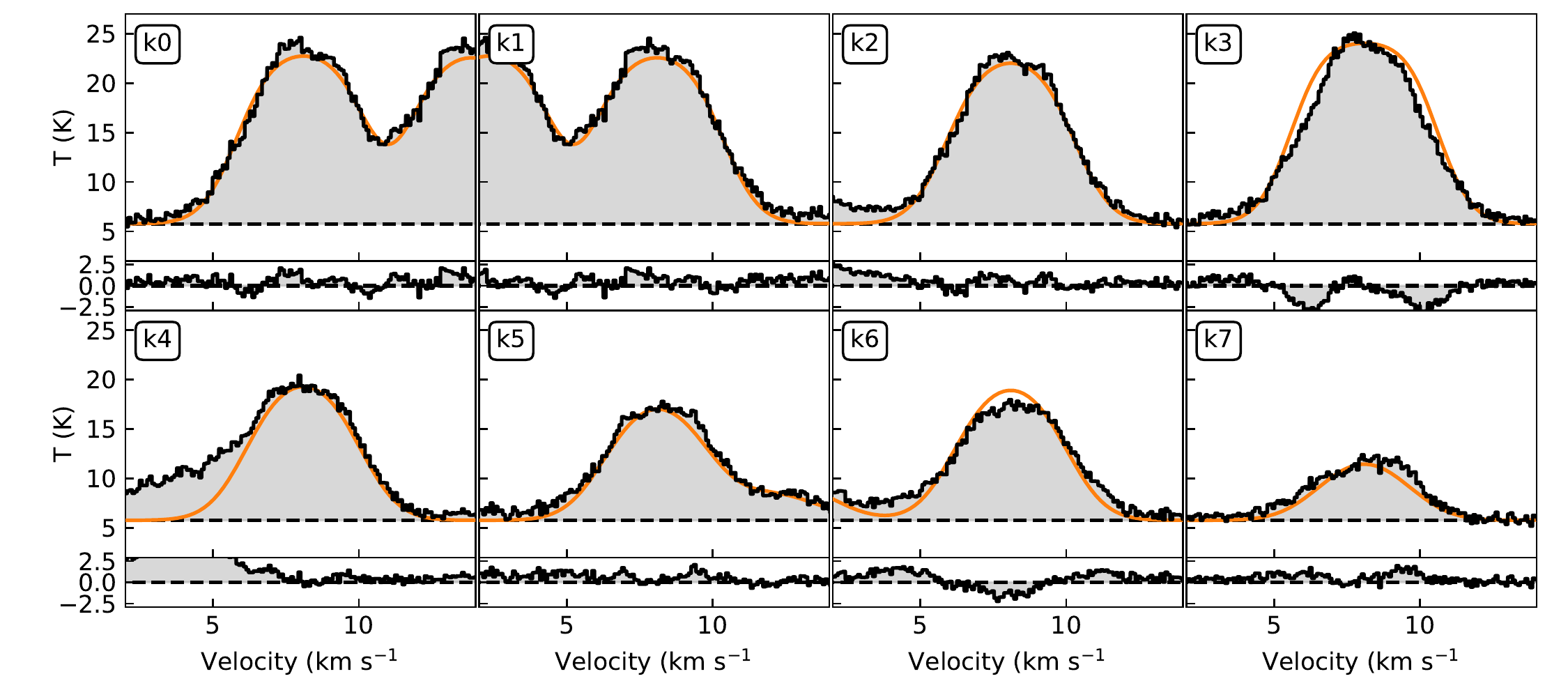}
\end{center}
\caption{Zoom-in spectra for each K-ladder component of \ce{CH3CN} J$=$12-11. The orange curve shows the best-fit 1-component LTE model. The bottom frame in each panel shows the residual from the fitting.}
\label{fig:asym}
\end{figure*}

\begin{figure*}
\begin{center}
\includegraphics[width=0.97\textwidth]{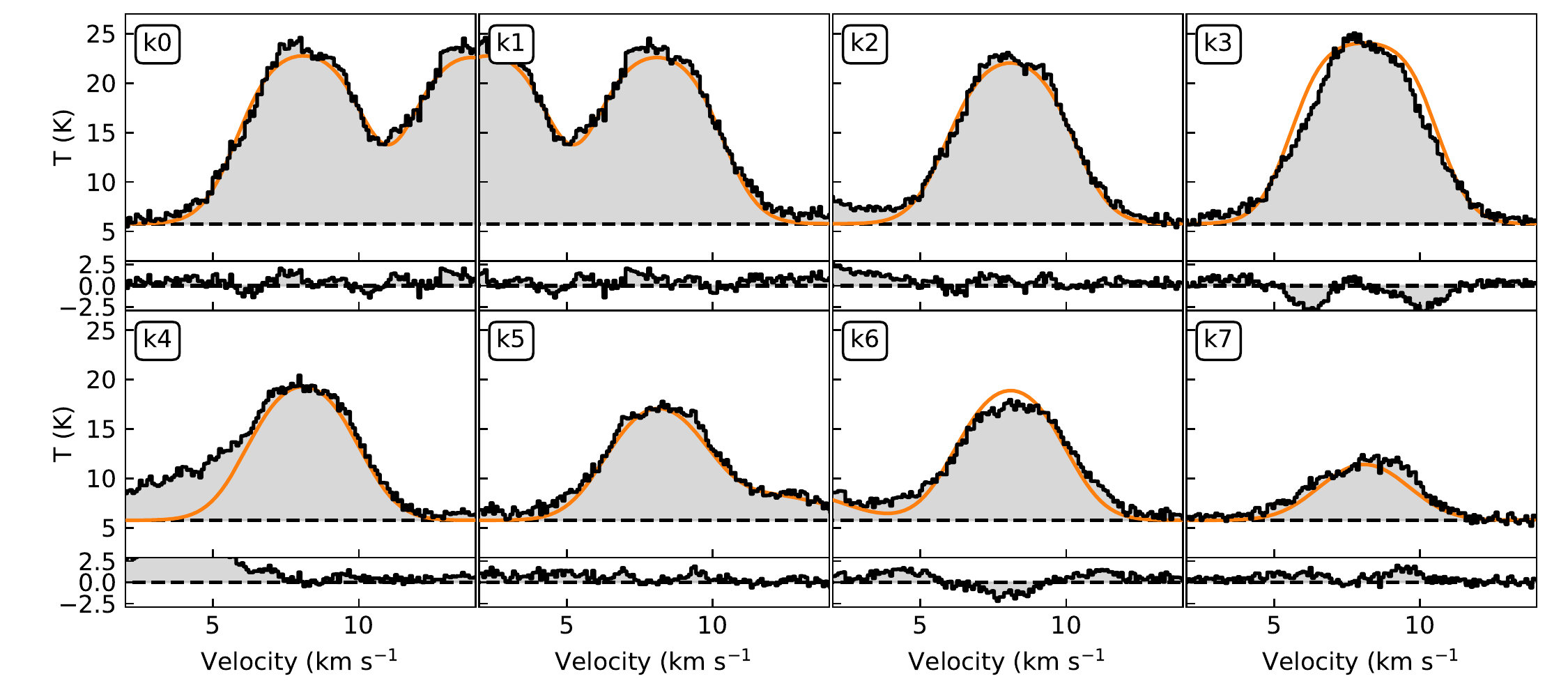}
\end{center}
\caption{Same as Figure \ref{fig:asym} but with model from LTE two species (Table \ref{tab:cassis}).}
\label{fig:asym2}
\end{figure*}

\begin{figure*}
\begin{center}
\includegraphics[width=0.97\textwidth]{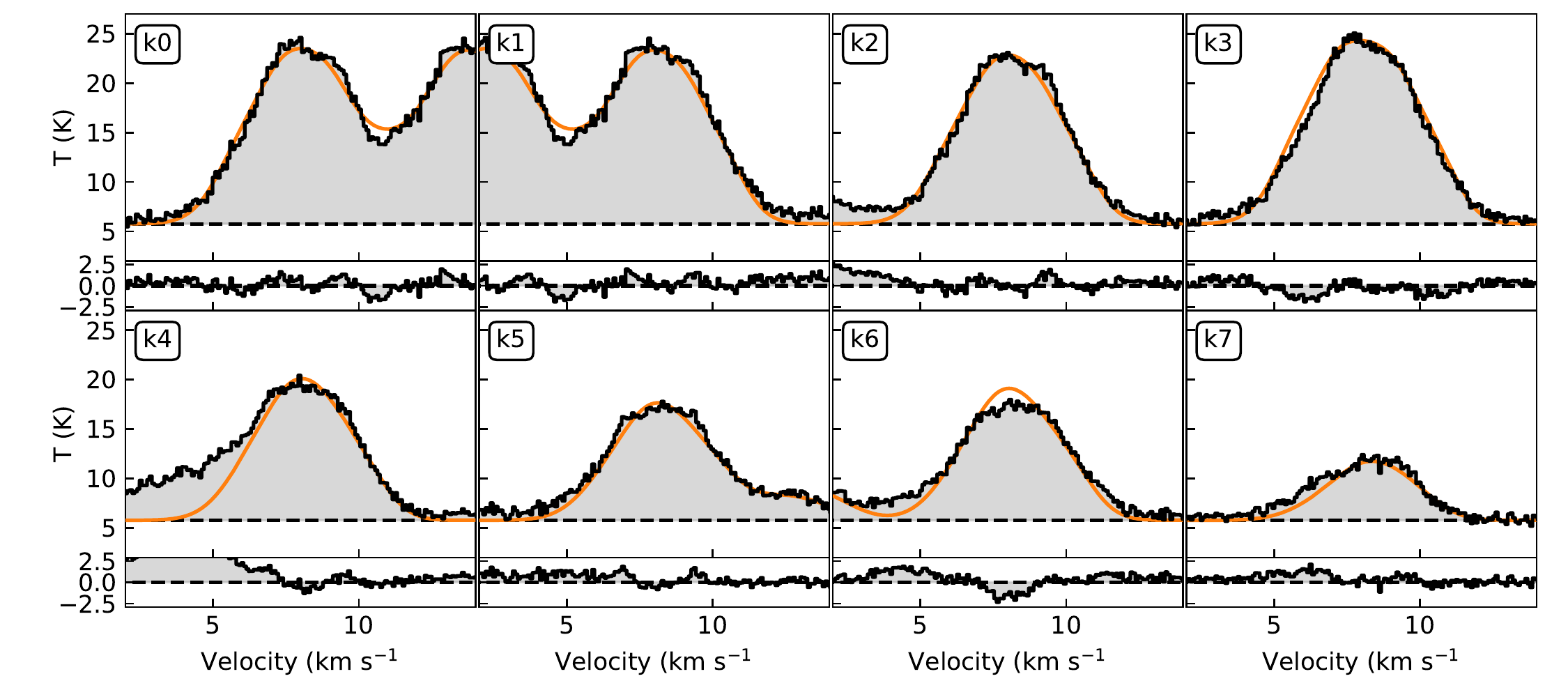}
\end{center}
\caption{Same as Figure \ref{fig:asym} but with model from LTE 2-comp (Table \ref{tab:cassis}).}
\label{fig:asym3}
\end{figure*}

\begin{figure*}
\begin{center}
\includegraphics[width=0.97\textwidth]{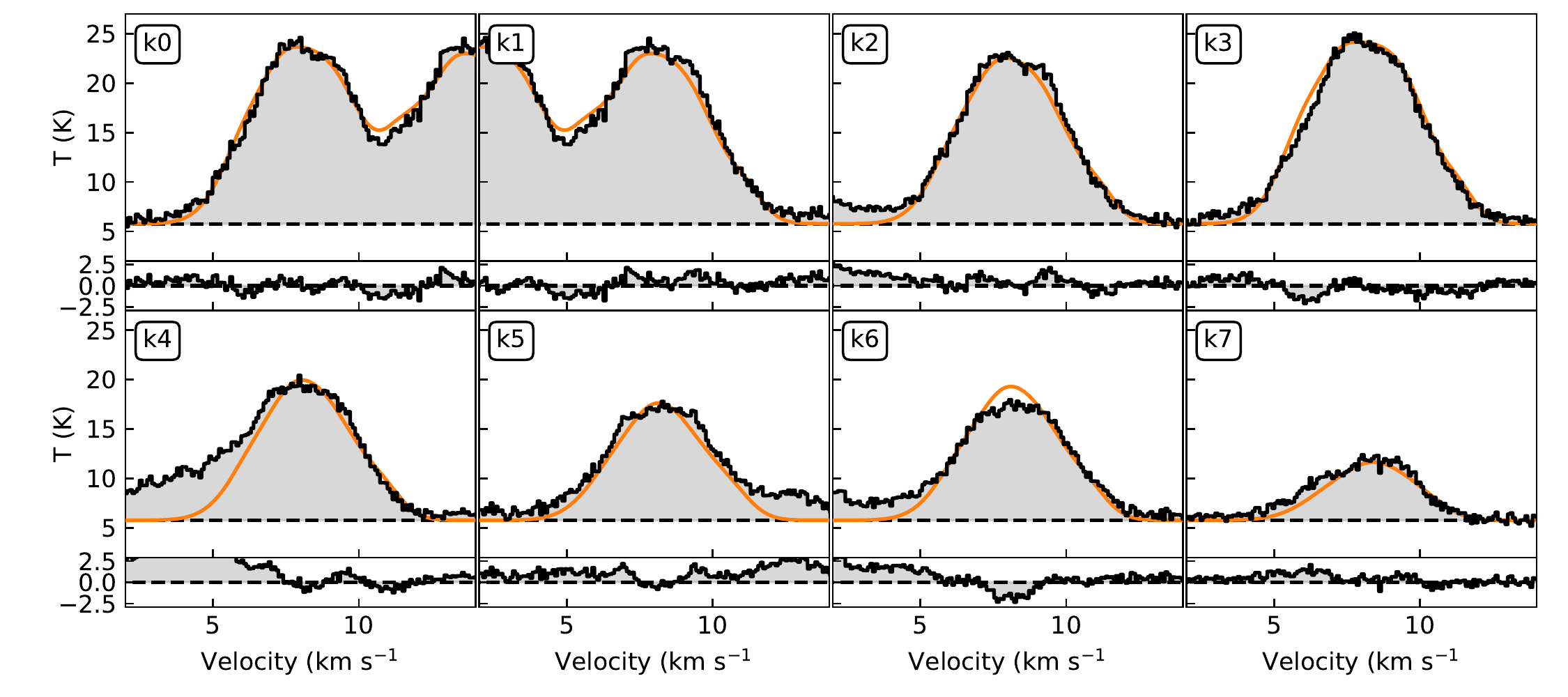}
\end{center}
\caption{Same as Figure \ref{fig:asym} but with model from non-LTE 2-comp (Table \ref{tab:cassis}).}
\label{fig:asym4}
\end{figure*}

\begin{figure*}
\begin{center}
\includegraphics[width=0.97\textwidth]{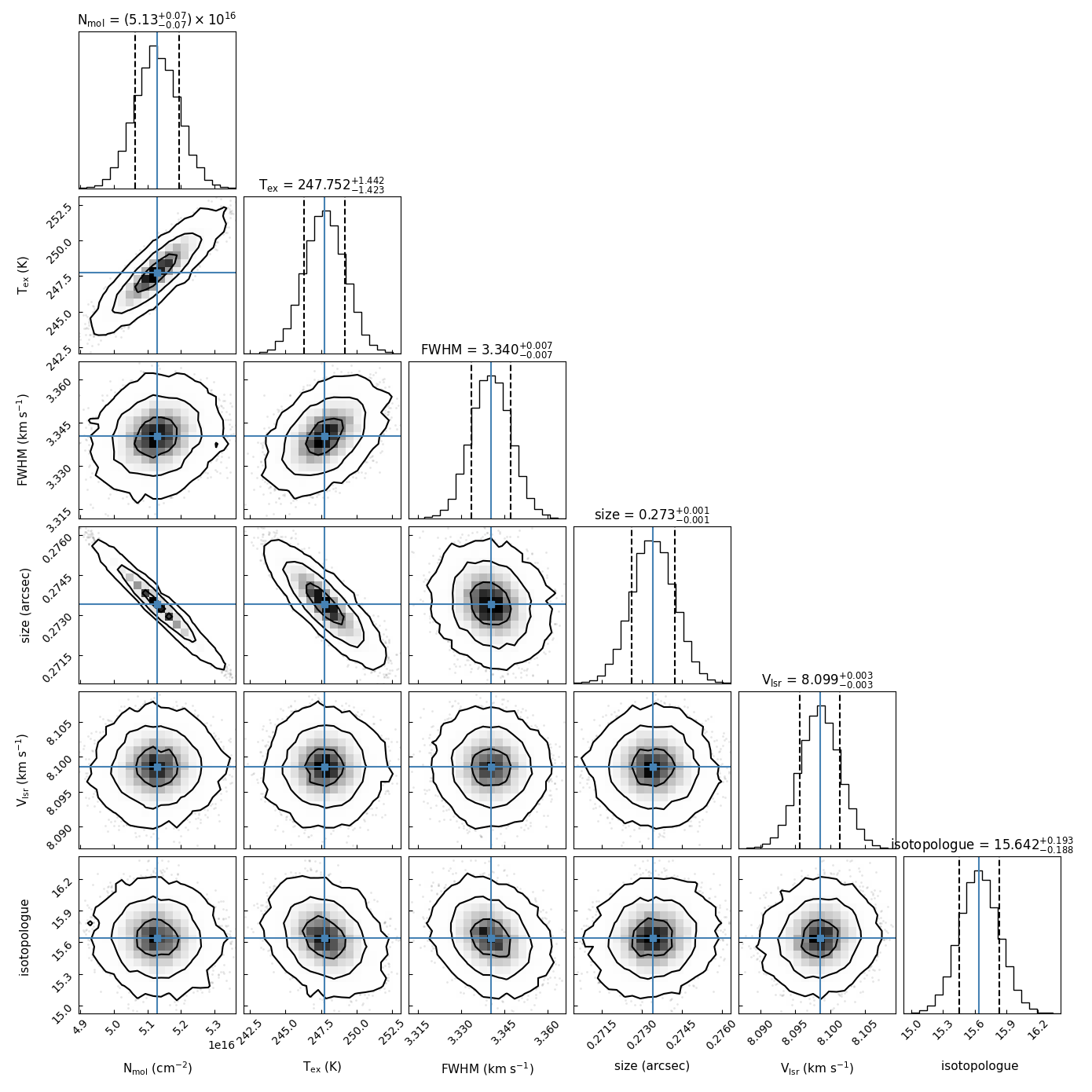}
\end{center}
\caption{Corner plot of the one-component MCMC fitting. The parameters from left to right are \ce{CH3CN} column density, excitation temperature, FWHM linewidth, source size, systemic velocity, and isotopologue ratio $\frac{\ce{CH3CN}}{\ce{CH3{}^13CN}}$.
}
\label{fig:corner_lte}
\end{figure*}

We conduct MCMC fitting to the \ce{CH3CN} and \ce{CH3{}^13CN} spectrum using CASSIS \citep{va15}. The information of the Markov chain for each model is listed in Table \ref{tab:mcmc}. 
The same process is applied for each of these four models, and here we show the plots for the LTE one-velocity component model as an example.
Long Markov chains are taken in order to better sample the parameter space. 
We use trace plot and autocorrelation time to test the convergence (Figures \ref{fig:trace} and \ref{fig:acf}); however, we note that these tests are only heuristic indicators of convergence \citep{ho18}. 
The $\chi^2$ and acceptance rate as functions of step are plotted with trace plots. The acceptance rate is from the number of the accepted and rejected proposals from the random walker (or Markov chain).
The recommanded value is 0.2--0.5 and 0.234 for best-performance in high-dimensional problems \citep{ho18}.
The autocorrelation time at lags is estimated using the Python Package statsmodels\footnote{https://www.statsmodels.org/stable/index.html}. 
We compare the estimated autocorrelation time $\tau_{\rm estimate}$ with different sample sizes. The autocorrelation time indicates a sequence within which the sampling points are similar. In other words, two sampled points with a separation $>\tau_{\rm estimate}$ are considered to be independent.
We find that $\tau_{\rm estimate}$ had become stable in larger sample sizes in our chain (Figure \ref{fig:acf}).
The trace plots and autocorrelation time also act as a guide to tune the steps of the walkers for each parameter. 
The step sizes are thus tuned to be comparable to the dispersion of the sample, and keep an acceptance rate between 0.2-0.5.
For each model, this procedure is repeated several times until a proper tuning is found.
We check the trace plots (as well as the $\chi^2$ derived by CASSIS), and remove the beginning of the chain (burn-in phase; 1500-15000 steps depending on models). To obtain independent points from the sample, we take one data point every $\tau_{\rm int}/2$ sample, i.e., thinning process; while the $\tau_{\rm int}$ can be very different among the fitting parameters (Table \ref{tab:mcmc}), here we take 1/2 of the smallest $\tau_{\rm int}$ in order to preserve a large sample (Table \ref{tab:cassis}).
The physical parameters and their uncertainties are thus estimated using these thinning samples.
Finally, the best fit of the four models is shown in Figures \ref{fig:asym} to \ref{fig:asym4}. The opacities of the \ce{CH3CN} K-ladder transitions for each component from the four best-fit models are listed in Table \ref{tab:tau}. The corner plot for the one velocity component model is shown in Figure \ref{fig:corner_lte}

\begin{table*}
    \centering
    \caption{MCMC fitting}
    \begin{tabular}{ccccc}
    \hline\hline
Model
& draws		
& chain length
& $\tau_{\rm int}$
& thinned sample \\

\hline
LTE model & 10$^6$ & 390513 & 45-398 & 17737 \\
\hline
LTE split iso. & 10$^6$  & 294053 & 62-2057 & 9465 \\
\hline
LTE 2-comp  & $2\times10^6$ &  397485 & 204-1191 & 3888 \\
\hline
non-LTE 2-comp  & 10$^6$ & 234956 & 466-1665 & 1001 \\
\hline

    \end{tabular}
    \label{tab:mcmc}
\end{table*}

%
\begin{table*}
    \centering
    \caption{Opacities of best-fit \ce{CH3CN} model
    \label{tab:tau}}
    \begin{tabular}{ccccccccccccc}
    \hline\hline
Model   
& component
& K0
& K1
& K2
& K3
& K4
& K5
& K6
& K7
\\
\hline
LTE model & - & 2.21 & 2.13 & 1.92 & 3.20 & 1.24 & 0.89 & 1.18 & 0.36\\
\hline
LTE two species & \ce{CH3CN} & & 2.23 & 2.15 & 1.93 & 3.21 & 1.24 & 0.89 & 1.17 & 0.35 \\
& \ce{CH3{}^13CN} & 0.21 & 0.20 & 0.17 & 0.27 & 0.09 & 0.06 & - & -\\
\hline
LTE 2-comp&
extend & 2.73 & 2.61 & 2.28 & 3.64 & 1.33 & 0.88 & 1.06 & 0.28\\
& compact & 4.32 & 4.18 & 3.78 & 6.41 & 2.54 & 1.87 & 2.55 & 0.80\\
\hline
non-LTE 2-comp&
extend & 2.82 & 2.51 & 2.30 & 3.75 & 1.45 & 0.96 & 1.41 & 0.33\\
& compact & 13.35 & 11.79 & 10.52 & 1.73 & 6.07 & 3.88 & 5.10 & 1.22\\ 
\hline
    \end{tabular}
\end{table*}


\section{Kinematic models}
\label{sec:rot}
\begin{figure*}
\begin{center}
\includegraphics[width=0.85\textwidth]{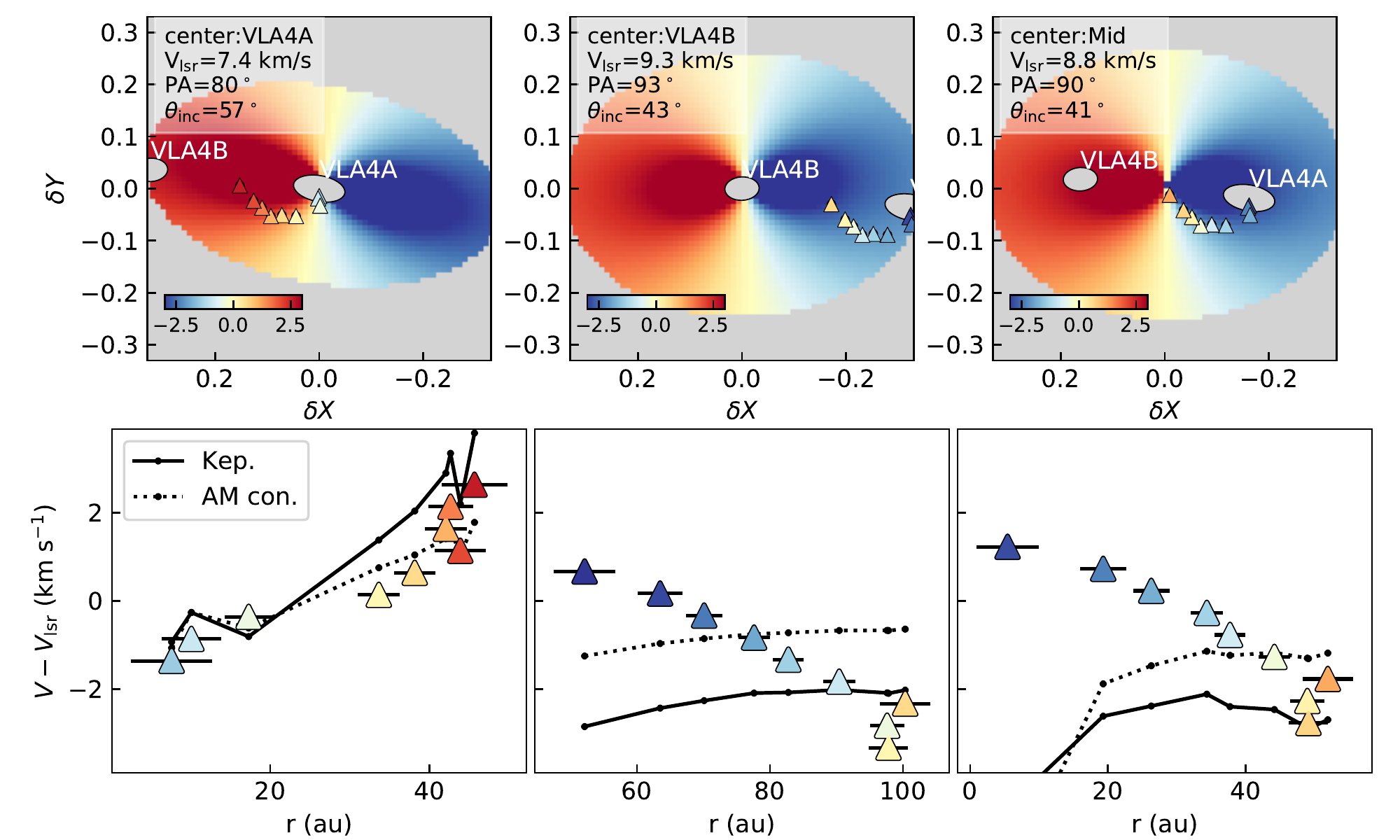}
\end{center}
\caption{
({\it top}) Gaussian centers of \ce{CH3CN} $12_3-11_3$ emission at different velocities from the uv-domain fitting in comparison with the line-of-sight velocity of a Keplerian disk model ($V\propto {\rm r}^{-0.5}$).
These models adopte three sets of $V_{\rm lsr}$, PA, $\theta_{\rm inc}$ listed in the upper-left corner from \citet{di21} for VLA4A/B circumstellar disks and the circumbinary disk.
({\it bottom})
Line-of-sight velocities of these points as a function of the inclination-corrected radius.
The solid and dotted lines show modeled velocities at the corresponding emitting positions assuming Keplerian motion ($V\propto {\rm r}^{-0.5}$) and conservation of angular momentum ($V\propto {\rm r}^{-1}$), respectively. Note that the modeled curves will only shift vertically if the velocity of the model is scaled.
}
\label{fig:rot_curve}
\end{figure*}

In order to probe the origin of the \ce{CH3CN} emission in SVS13A, we compare kinematic models with the three disks from \citet{di21}, i.e., the circumstellar disks of VLA4A, VLA4B as well as the circumbinary disk (Figure \ref{fig:rot_curve}). The emitting position at each velocity is obtained by Gaussian fitting in uv-space (Figure \ref{fig:uv_chan}). 
Using the central position, position angle, and inclination angle of the VLA4A, VLA4B disks, and the circumbinary disk from \citet{di21}, we derive the de-projected radius of the \ce{CH3CN} emission from the Gaussian position.
Then, we plot the measured and modeled velocities of the \ce{CH3CN} emission corresponding to these three disks. 
Comparing the line-of-sight velocity of a Keplerian model, the \ce{CH3CN} emission seems to trace part of a disk of VLA4A (see the top panel of Figure \ref{fig:rot_curve}). However, at least at the current spatial resolution, we find that the rotation curve cannot be simply interpreted by a simple slope for Keplerian motion $V\propto r^{-0.5}$ or conservation of angular momentum (infalling disk) $V\propto r^{-1}$ (see Figure \ref{fig:rot_curve}).

\section{Channel maps}
\label{sec:dcn_chan}
\begin{figure*}
\begin{center}
\includegraphics[width=0.99\textwidth]{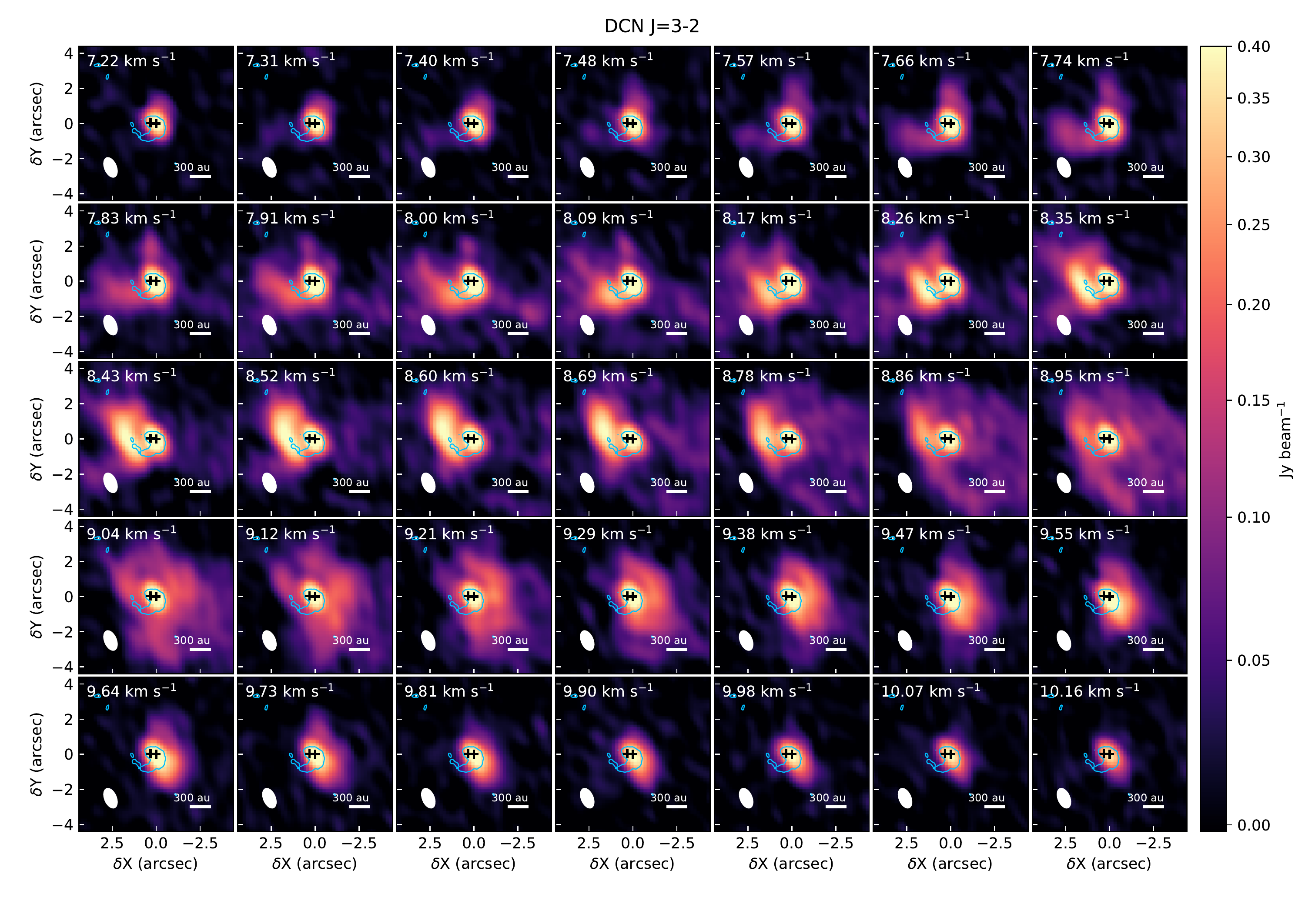}
\end{center}
\caption{DCN (3-2) channel map. The blue contour shows the ALMA continuum emission at 3$\sigma$.
}
\label{fig:dcn_chan}
\end{figure*}

Figure \ref{fig:dcn_chan} shows the channel maps from 7.22 to 10.16 km s$^{-1}$. 
The full channel map shows that, in addition to the infalling streamer, DCN actually traces more extended gas components.
At low velocities $\sim7.4-8.0$ km s$^{-1}$, DCN traces a gas component offset from the continuum streamer, which is also seen in the \ce{C^{18}O} emission (Figure \ref{fig:c18o_chan}). At $\sim8.26-8.95$ km s$^{-1}$, the DCN emission is dominated by the infalling streamer. The red-shifted emission $>$9 km s$^{-1}$ is likely associated with another extended component.

\begin{figure*}
\begin{center}
\includegraphics[width=0.99\textwidth]{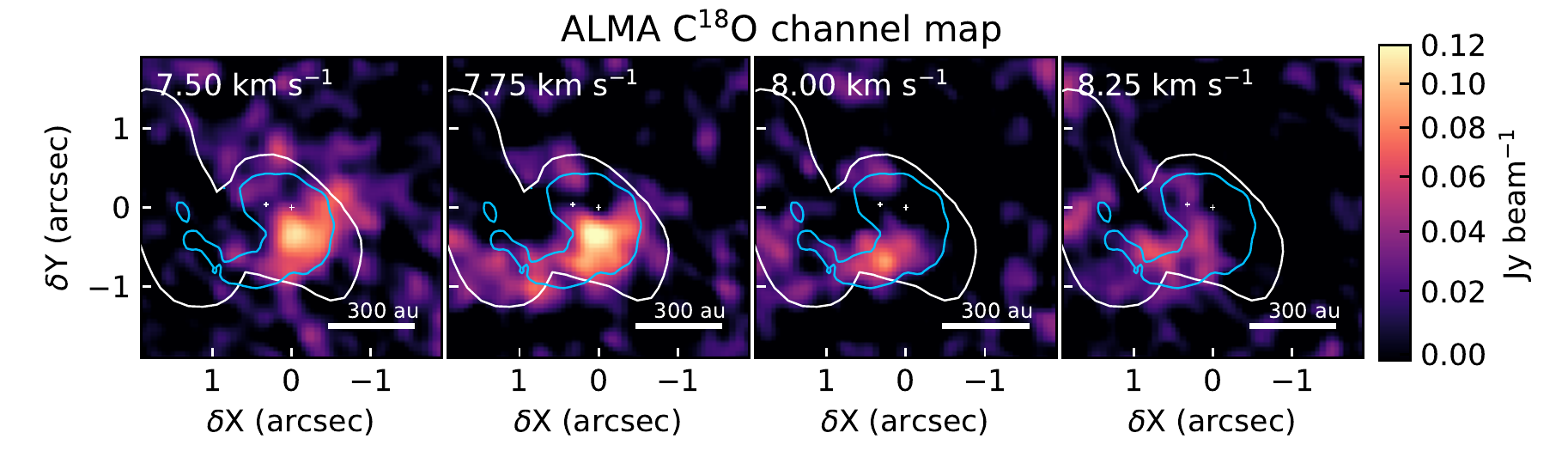}
\end{center}
\caption{ALMA \ce{C^{18}O} channel map. The blue contour shows the ALMA continuum emission at 3$\sigma$ and the white contour shows the DCN (3-2) peak intensity map at 25$\sigma$.
}
\label{fig:alma_c18o}
\end{figure*}

The \ce{C^{18}O} blue-shifted component is found to spatially match the inner part the spiral/streamer seen in the continuum emission \citep{to18}.
Figure \ref{fig:alma_c18o} shows the channel maps of this blue-shifted component. 
The \ce{C^18O} emission at this velocity range $7.50-8.25$ km s$^{-1}$ is consistent with the velocity map of the DCN blue-shifted component (Figure \ref{fig:decom}), and spatially well matches the continuum spiral.

\section{Decomposition of DCN (3-2) spectra}
\label{sec:decom}
\begin{figure*}
\begin{center}
\includegraphics[width=0.99\textwidth]{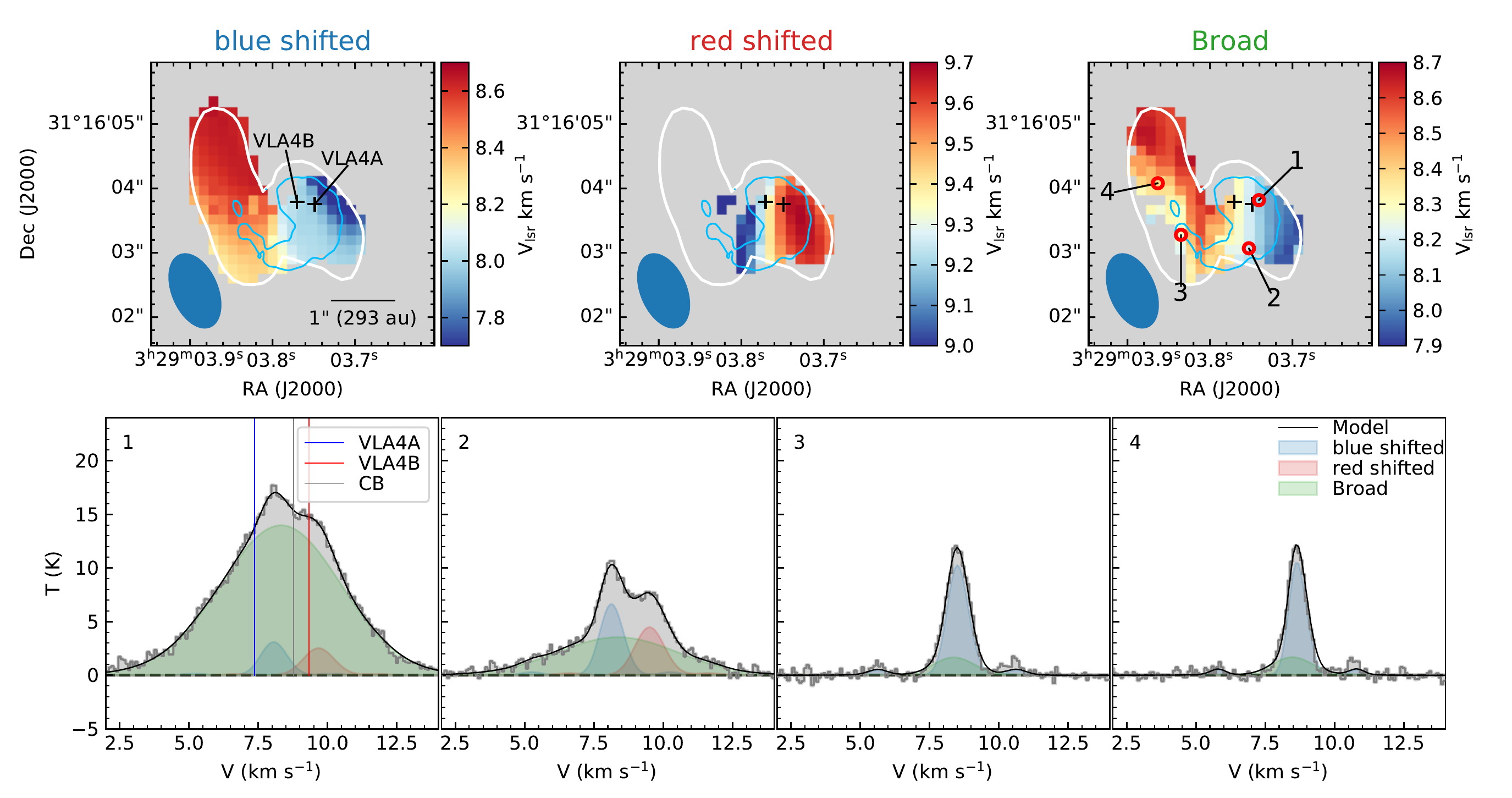}
\end{center}
\caption{
({\it Top}) DCN (3-2) centroid-velocity maps of the three velocity components from the hyperfine structure fitting. The white contour shows the peak intensity above 25$\sigma$. The blue contour shows the ALMA 1.3 mm continuum emission at the 3$\sigma$ level. ({\it Bottom}) The spectra of DCN (3-2)
with the hyperfine structure fitting toward the four positions marked in the top right panel. The three vertical lines represent the systemic velocites of VLA4A, VLA4B, and the circumbinary disk from \citet{di21}.
}
\label{fig:decom}
\end{figure*}

We conduct HFS fitting in the DCN (3-2) channel map. Three different velocity components are found (Figure \ref{fig:decom}). 
The fitting is firstly conducted with one, two, and three velocity components.
To proceed with the fitting, we took several boundary constraints in the three components; blue-shifted component: $V_{\rm lsr}<8.8$ km s$^{-1}$ and $\sigma <0.6$ km s$^{-1}$, red-shifted component: $V_{\rm lsr}>8.8$ km s$^{-1}$ and $\sigma <0.6$ km s$^{-1}$, and broad component: $\sigma <2$ km s$^{-1}$.

For each pixel, we select the model with one, two or three velocity components based on Akaike Information Criterion (AIC),
\begin{equation}
    AIC=2{\rm k}+\chi^2+C
\end{equation}
where k is the degree of freedom (3, 6, and 9), $\chi^2$ is obtained from the fitting residual, and C is a constant (see Appendix E. in \citealt{ch20}).
Comparing the AIC in each pixel, we use the model with lowest AIC value.

\end{document}